\documentclass[11pt,a4paper]{article}

\usepackage{jheppub}
\usepackage{graphicx}
\usepackage{dcolumn}
\usepackage{bm}
\usepackage{amsmath}
\usepackage{braket}
\usepackage{slashed}    
\usepackage{epstopdf}
\usepackage{placeins}
\usepackage{multirow}
\usepackage{makecell}
\usepackage{soul}
\usepackage{braket}
\usepackage[shortlabels]{enumitem}
\usepackage{placeins}
\usepackage{cleveref}
\usepackage{subcaption}
\usepackage[font=small, labelfont=bf, textfont={small,it}]{caption}
\usepackage{xcolor}

\graphicspath{{figs/}}
\renewcommand\thesection{\arabic{section}}
\newcommand{\beq}{\begin{eqnarray}}
\newcommand{\eeq}{\end{eqnarray}}
\newcommand{\beqnn}{\begin{eqnarray*}}
\newcommand{\eeqnn}{\end{eqnarray*}}

\newcommand{\Var}{\ensuremath{\mathrm{Var}}}
\newcommand{\CP}{\mathrm{CP}}
\newcommand{\SU}{\mathrm{SU}}

\newcommand{\dd}{{\mathrm{d}}}
\newcommand{\nstep}{n_{\mathrm{step}}}
\newcommand{\nev}{n_{\mathrm{ev}}}
\newcommand{\nstat}{n_{\mathrm{stat}}}
\newcommand{\nrelax}{n_{\mathrm{relax}}}
\newcommand{\tauint}{\tau_{\mathrm{int}}}

\newcommand{\DKL}{\tilde{D}_{\mathrm{KL}}}
\newcommand{\ESS}{\mathrm{ESS}}
\newcommand{\NE}{\mathrm{NE}}

\begin{document}

\title{Mitigating topological freezing using out-of-equilibrium simulations}

\author[a]{Claudio Bonanno,}
\affiliation[a]{Instituto de F\'isica T\'eorica UAM-CSIC, c/ Nicol\'as Cabrera 13-15, Universidad Aut\'onoma de Madrid, Cantoblanco, E-28049 Madrid, Spain}
\emailAdd{claudio.bonanno@csic.es}

\author[b]{Alessandro Nada,}
\affiliation[b]{Dipartimento di Fisica,  Universit\'a degli Studi di Torino and INFN, Sezione di Torino, Via Pietro Giuria 1, I-10125 Turin, Italy}
\emailAdd{alessandro.nada@unito.it}

\author[c]{Davide Vadacchino}
\affiliation[c]{Centre for Mathematical Sciences, University of Plymouth, Plymouth, PL4 8AA, United Kingdom}
\emailAdd{davide.vadacchino@plymouth.ac.uk}

\abstract{Motivated by the recently-established connection between Jarzynski's equality and the theoretical framework of Stochastic Normalizing Flows, we investigate a protocol relying on out-of-equilibrium lattice Monte Carlo simulations to mitigate the infamous computational problem of topological freezing. We test our proposal on $2d$ $\mathrm{CP}^{N-1}$ models and compare our results with those obtained adopting the Parallel Tempering on Boundary Conditions proposed by M.~Hasenbusch, obtaining comparable performances. Our work thus sets the stage for future applications combining our Monte Carlo setup with machine learning techniques.}


\maketitle
\flushbottom

\section{Introduction}\label{sec:intro}

It is well known that Markov Chain Monte Carlo simulations of 
lattice gauge theories based on
local updating algorithms are affected by critical slowing down.
The integrated auto-correlation time $\tau$ associated with a given lattice observable $\mathcal{O}$ diverges as the continuum limit is approached, in a way that is naturally described in terms of the correlation length $\xi$ of the system. 

While for lattice observables like the elementary plaquette 
or the Polyakov line $\tau$ diverges as a power law, 
$\tau\sim\xi^z$, with $z$ typically of order 2 or less, 
for topological quantities such as the topological 
charge $Q$ this divergence is much more dramatic, 
see Refs.~\cite{Alles:1996vn,deForcrand:1997yw,Lucini:2001ej,
Leinweber:2003sj,Lucini:2004yh,DelDebbio:2004xh, 
DelDebbio:2006yuf, Schaefer:2010hu, Luscher:2011kk,
Laio:2015era, Flynn:2015uma,Bonati:2016tvi, Hasenbusch:2017unr,
Bonati:2017woi,Bonanno:2018xtd,Bonanno:2020hht,
Athenodorou:2021qvs,Bennett:2022ftz}. There is by now vast numerical 
evidence that both in $4d$ $\SU(N)$ gauge theories and in 
two-dimensional models, the divergence is exponential, 
$\tau(Q)\sim e^{\xi}$, see Refs.~\cite{Lucini:2004yh,
DelDebbio:2004xh, DelDebbio:2006yuf, Bonanno:2018xtd}.

Topological critical slowing down can be understood as follows.
Sufficiently close to the continuum limit, a proper definition of 
topological charge is recovered for lattice gauge field configurations. 
This definition can be used to partition configuration space 
into sectors, separated by free energy barriers. The height 
of these barriers diverges with $\xi$ and, as the continuum limit is 
approached, the disconnected topological sectors of the continuum theory emerge. The growth of the energy barriers also suppresses the tunneling
rate of a Markov Chain generated by a local updating algorithm, that
thus remains trapped in one fixed topological sector. This effect, known 
as \emph{topological freezing}, leads to a loss of ergodicity 
and introduces large systematic effects, especially in the calculation
of topological observables such as the topological susceptibility. 

Algorithms that exactly solve the problem of topological 
freezing are only known for specific low-dimensional 
toy models, see for example Ref.~\cite{Bonati:2017woi}. Yet, several suggestions have 
been made in the last few years to make the growth of 
$\tau$ for topological observables milder, namely from exponential 
to polynomial. Examples are simulations with 
Open Boundary Conditions (OBCs)~\cite{Luscher:2011kk,Luscher:2012av}, 
Parallel Tempering on Boundary Conditions 
(PTBC)~\cite{Hasenbusch:2017unr,Bonanno:2020hht},
metadynamics~\cite{Laio:2015era,Eichhorn:2023uge}, 
density of states 
methods~\cite{Cossu:2021bgn,Borsanyi:2021gqg}, master 
field simulations~\cite{Luscher:2017cjh}, machine-learning based 
approaches such as Normalizing Flows~\cite{Kanwar:2020xzo,Nicoli:2020njz,papamakarios2021,Abbott:2023thq}, 
and many others~\cite{Bietenholz:2015rsa,Funcke:2019zna,Albandea:2021lvl,Boyle:2024nlh}.

In this study, we propose a novel algorithm aimed at mitigating
topological freezing. As our strategy builds on features of simulations
with OBCs as well as on the PTBC algorithm proposed by 
M.~Hasenbusch in Ref.~\cite{Hasenbusch:2017unr}, we briefly summarize their main properties below.

In theories defined on a continuum space-time, topological charge 
is integer valued as a consequence of Periodic Boundary 
Conditions (PBCs) in time.
Abandonding PBCs in favour of OBCs corresponds to eliminating the
barriers between topological sectors, which are not disconnected anymore.
Performing Monte Carlo simulations with OBCs thus allows the Markov Chain to 
switch topological sector more easily, resulting in a dramatic reduction 
of topological auto-correlation time. In particular, it was shown 
in Ref.~\cite{Luscher:2011kk,Luscher:2012av} that the divergence of 
$\tau$ is reduced to a polynomial one, $\tau\sim \xi^2$.
Yet, this approach suffers from some drawbacks. 
As a matter of fact, OBCs introduce nonphysical effects, which have to 
be avoided by only computing correlation functions in the bulk 
of the lattice. Hence, larger volumes are required in order 
to keep finite size effects under control while still avoiding 
unwanted systematic errors. In this respect, the PTBC algorithm, proposed for $2d$ $\CP^{N-1}$ models in Ref.~\cite{Hasenbusch:2017unr} and recently implemented also for $4d$ $\SU(N)$ gauge theories in Ref.~\cite{Bonanno:2020hht} aims at having the best of both worlds. By combining PBCs and OBCs simulations in the framework of the parallel tempering
idea, it exploits the improved scaling of the auto-correlation time 
of the topological charge in systems with OBCs simulations while, at the same time, avoiding the drawbacks metioned above, as physical quantities are computed with PBCs. 
The PTBC algorithm has been recently employed in a variety of cases, 
demonstrating a dramatic reduction of the auto-correlation time of 
the topological charge, and improving state-of-the-art results for 
several topological and non-topological 
quantities~\cite{Berni:2019bch,Bonanno:2020hht,
Bonanno:2022yjr,Bonanno:2022hmz,DasilvaGolan:2023cjw,Bonanno:2023hhp,Bonanno:2024ggk}.

The strategy proposed in this study shares its roots with 
the PTBC algorithm, as it still combines the use of Open and Periodic 
boundary conditions. However, it makes use of 
\emph{out-of-equilibrium evolutions} based on the 
well-known Jarzynski's equality, see Ref.~\cite{Jarzynski:1996oqb}.
A fundamental result of non-equilibrium statistical mechanics,
Jarzynski's equality has been extensively used in recent years
in several contexts in lattice field theories, ranging from
the computation of interface free energies, see Ref.~\cite{Caselle:2016wsw}, 
of the QCD equation of state, see Ref.~\cite{Caselle:2018kap}, of 
renormalized coupling of $\SU(N)$ gauge theories, 
see Ref.~\cite{Francesconi:2020fgi}, and 
in the study of the entanglement entropy from the lattice, 
see Ref.~\cite{Bulgarelli:2023ofi}.

The underlying idea is to sample the configuration space of 
the system with PBCs 
using a previous sampling of the configuration space of the
system with OBCs, the latter being used as a starting 
point for non-equilibrium evolutions.
Using Jarzynski's equality, the expectation values of the desired 
observables can then be obtained for the system with PBCs 
through a reweighting procedure. 
In this approach, the decorrelation of the topological charge still benefits from the presence of OBCs, while avoiding their pitfalls, 
with a cost overhead that will be quantified by the length of the out-of-equilibrium trajectory.

The main motivation behind the present study is a 
recent development in the field of machine-learning based 
on the combination of Jarzynski's equality with Normalizing 
Flows (NFs).
It this new framework, known as Stochastic Normalizing Flows (SNFs), 
see Refs.~\cite{wu2020stochastic,Caselle:2022acb}, out-of-equilibrium 
Monte Carlo evolutions are combined with discrete coupling layers
(the same building blocks composing NFs) to achieve a 
substantial improvement of sampling efficiency with respect to a 
purely stochastic approach. As our strategy is rooted on Jarzynski's 
equality too, SNFs would be a natural future direction for
the application of our proposal.

The aim of the present study is thus to probe the use of the
out-of-equilibrium methods in the computation of the topological
observables in order to set the stage for this future development. Its technical feasibility will be explored, and 
its performance compared with that of the PTBC algorithm. 
We will focus on the $2d$ $\CP^{N-1}$ models, see Refs.~\cite{DAdda:1978vbw, Luscher:1978qe, Vicari:2008jw,Shifman:2012zz}, which are a very 
popular test bed for new numerical approaches, see Refs.~\cite{Campostrini:1988cy,
Campostrini:1992ar,Campostrini:1992it,Alles:1997nu,
DelDebbio:2004xh,Bietenholz:2010xg,Hasenbusch:2017unr,
Bonanno:2018xtd,Berni:2019bch,Berni:2020ebn,Bonanno:2022dru}. They are
generally simpler to study on the lattice compared to QCD while still
exhibiting non-trivial topological features, which they share 
with $4d$ $\SU(N)$ gauge theories. 
A preliminary version of the results discussed in this manuscript was 
presented at the 2023 Lattice Conference, and can be found in Ref.~\cite{Bonanno:2023ier}.

This paper is organized as follows. In Section~\ref{sec:setup} we introduce
the numerical setup used in this study, with a heavy focus on the features of out-of-equilibrium evolutions. 
The numerical results obtained with this method are presented
and discussed in Section~\ref{sec:res}. Finally, in 
Section~\ref{sec:conclu} we draw our conclusions and point to the 
future directions of this investigation.

\section{Numerical setup}\label{sec:setup}

In this section, the numerical setup used in this study is introduced.
The $2d$ $\CP^{N-1}$ models are defined, along with the local updating
algorithm used to generate their configurations. Out-of-equilibrium
methods are also introduced, along with a detailed explanation on how they are used to carry out the
program sketched above.

\subsection{Lattice discretization of $2d$ $\CP^{N-1}$ models}

Consider the following continuum Euclidean action of $2d$ $\CP^{N-1}$ models~\cite{DAdda:1978vbw,Luscher:1978qe}:
\beq\label{eq:continuum_action}
S[\overline{z},z,A]=\int d^2x\left[ \frac{N}{g}\overline{D}_\mu \overline{z}(x) D_\mu z(x) \right],
\eeq
where $g$ is the 't Hooft coupling, 
$z=(z_1,\dots,z_N)$ is a $N$-component complex scalar 
satisfying $\overline{z}z = 1$, and 
$D_\mu \equiv \partial_\mu + i A_\mu$ is the $\mathrm{U}(1)$ covariant
derivative, where $A_\mu(x)$ is a non-propagating $\mathrm{U}(1)$ gauge
field\footnote{The non-propagating gauge field $A_\mu(x)$ could be 
integrated out and expressed in terms of
$z(x)$~\cite{Luscher:1978qe,Vicari:2008jw}. However, this 
formulation is more convenient for the purpose of lattice simulations.}. 
An integer-valued topological charge can be conveniently expressed 
in terms of the gauge field according to Refs.~\cite{DAdda:1978vbw,Luscher:1978qe} as follows,
\beq\label{eq:continuum_topcharge}
Q=\frac{1}{2\pi} \epsilon_{\mu\nu} \int d^2x \, \partial_\mu A_\nu(x) \in \mathbb{Z},
\eeq
while the topological susceptibility, which is our 
main observable of interest, is defined as usual as (here $V$ is the space-time volume):
\beq
\chi = \lim_{V\to\infty} \frac{\braket{Q^2}}{V}.
\eeq

\begin{figure}[!t]
\centering
\includegraphics[scale=0.95]{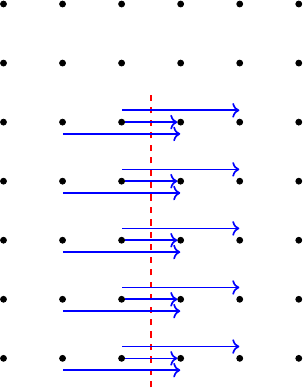}
\caption{Figure taken from Ref.~\cite{Berni:2019bch}. The dashed line represents the defect $D$, while the solid arrows represent links or product of links crossing the defect orthogonally, and thus getting suppressed by factor(s) of $c(n)$ according to the definition of $k_\mu^{(n)}(x)$ in Eq.~\eqref{eq:k_mu}. In this case the defect is placed on the space boundary (i.e., along the $\mu=1$ direction), so that only temporal links $U_0$ will cross it.}
\label{fig:defect}
\end{figure}

We discretize the action, Eq.~\eqref{eq:continuum_action}, on a 
square lattice of size $L$, using the $O(a)$ tree-level 
Symanzik-improved lattice action defined in 
Ref.~\cite{Campostrini:1992ar}. 
A crucial ingredient in this study is the choice of boundary conditions.
They are imposed as periodic in every direction and for every 
point of the boundary of the lattice, except for a segment 
of length $L_d$ along the spatial boundary. In the following, 
this segment will be known as
\emph{defect} and will be denoted by $D$. 
Along the defect we impose OBCs. Our purpose is to
gradually evolve these to PBCs using the out-of-equilibrium evolutions 
described in Sec.~\ref{sec:off_eq_evo}. 
This choice of boundary conditions can be encoded directly into the
dynamics by considering the following family of lattice actions, each
labeled by an integer $n$,
\beq\label{eq:lattice_action}
\begin{aligned}
S_L^{(n)}[\overline{z},z,U] = -2 N \beta\sum_{x,\mu} \Big\{ &k_\mu^{(n)}(x) c_1\Re\left[\overline{U}_\mu(x)\overline{z}(x+\hat{\mu})z(x)\right] + \\
&\\[-1em]
+\,&k_\mu^{(n)}(x+\hat{\mu})k_\mu^{(r)}(x)c_2\Re\left[\overline{U}_\mu(x+\hat{\mu})\overline{U}_\mu(x)\overline{z}(x+2\hat{\mu})z(x)\right]\Big\}~,
\end{aligned}
\eeq
where $\beta$ is the inverse bare 't Hooft coupling, 
$c_1=4/3$ and $c_2=-1/12$ are Symanzik-improvement coefficients, 
and $U_\mu(x) = \exp\{i A_\mu(x)\}$ are the $\mathrm{U}(1)$ gauge 
link variables. 
The factors $k_\mu^{(n)}(x)$ appearing in 
Eq.~\eqref{eq:lattice_action} are used to implement different 
boundary conditions for the link variables crossing the defect $D$.
They are defined as follows,
\beq\label{eq:k_mu}
k_\mu^{(n)}(x) \equiv
\begin{cases}
c(n)\, , &\quad x \in D \wedge \mu=0\, , \\
1\, , &\quad \text{otherwise},\\
\end{cases}
\eeq
with $0\le c(n)\le 1$ a function that interpolates between
$c=0$ and $c=1$ corresponding, respectively, to OBCs and PBCs.
The setup described above is sketched in Fig.~\ref{fig:defect}. 

The behaviour of the model defined above can be clarified as 
follows. When $c=1$, 
the factors $k_\mu^{(n)}(x)$ are everywhere equal to $1$. Then, the action
in Eq.~\eqref{eq:lattice_action} reduces to the standard 
Symanzik-improved action of $2d$ $\CP^{N-1}$ models with PBCs. 
When $0\le c < 1$, the coupling $\beta$ on the links crossing 
$D$ get suppressed by a factor of $c$ each. 
This corresponds to a suppression of the force entering 
the updating procedure for those specific links and of the corresponding
site variables, and to its vanishing when $c=0$.  This condition
realizes OBCs, as the vanishing of the force implies no update of the involved variable.
In this work we chose a simple linear interpolation between OBCs and PBCs, namely, $c(n)=n/\nstep$, with $c(n=0)=0$ (OBCs case) and $c(\nstep)=1$ (PBCs case).

For the updating procedure, we relied on the standard 4:1 combination of 
Over-Relaxation (OR) and over-Heat-Bath (HB) local update algorithms, see Ref.~\cite{Campostrini:1992ar} for more details. The full lattice 
update sweeps were supplemented with \emph{hierarchical} updates 
of sub-regions of the lattice centered on the defect. 
These allow to update the links and sites close to $D$,
where the creation/annihilation of new topological excitations occurs 
most likely, thereby improving the evolution of the 
Markov Chain between topological sectors. These hierarchical 
updates were designed along the same lines 
of Ref.~\cite{Hasenbusch:2017unr}, where more details may be found. Hierarchical updates were alternated with translations in randomly-chosen directions by one lattice spacing of the position of the defect on the periodic replica (which is translation-invariant), so that topological excitations are created/annihilated in different places around the lattice. Such translations are effectively achieved by simply translating the site/link variables of the periodic field configurations.

As discussed in the Introduction, the measurement of lattice 
out-of-equilibrium evolutions happens on systems with PBCs. 
In the case of PBCs, no unphysical effects coming from the fixed boundaries
are present and we can safely rely on several different 
discretizations of the global topological charge defined in 
Eq.~\eqref{eq:continuum_topcharge} for the $2d$ $\CP^{N-1}$ models on a torus. 
In this study, we will adopt the so called \emph{geometric} definition as in Ref.~\cite{Campostrini:1992ar}, which is known to be integer valued 
for every lattice configuration,
\beq\label{eq:geometric_lattice_charge}
Q_{\rm geo}[U] &=& \frac{1}{2\pi}\sum_{x} \Im \left\{ \log \left[\Pi_{01}(x)\right] \right\} \in \mathbb{Z},
\eeq
where $\Pi_{\mu\nu}(x) \equiv U_\mu(x)U_\nu(x+a\hat{\mu})\overline{U}_\mu(x+a\hat{\nu})\overline{U}_\nu(x)$
is the elementary plaquette.

The geometric lattice topological charge in Eq.~\eqref{eq:geometric_lattice_charge} is computed 
on configurations that have undergone a small amount of smoothing of the gauge 
and matter fields. This is done in order to remove the effects of ultraviolet (UV) 
fluctuations at the scale of the lattice spacing. 
Several different smoothing methods have been proposed in the literature, 
such as cooling, see Refs.~\cite{Berg:1981nw,Iwasaki:1983bv,Itoh:1984pr,
Teper:1985rb,Ilgenfritz:1985dz,Campostrini:1989dh,Alles:2000sc}, 
stout smearing, see Refs.~\cite{APE:1987ehd, Morningstar:2003gk} or 
gradient flow, see Refs.~\cite{Luscher:2009eq, Luscher:2010iy}. 
From the numerical standpoint, all these methods have been shown to give 
consistent results when properly matched to one another, see 
Refs.~\cite{Alles:2000sc, Bonati:2014tqa, Alexandrou:2015yba}. 
In this study, we adopt cooling for its simplicity and for its limited numerical
cost. A single cooling step consists in aligning, 
site by site and link by link, 
each variable $z(x)$ and $U_\mu(x)$ to their corresponding local forces. 
As shown in Ref.~\cite{Alexandrou:2015yba}, the specific action from which 
the local force is computed does not need to be the one used for the 
Monte Carlo evolution. Thus, we rely on the non-improved action, i.e., 
the action in Eq.~\eqref{eq:lattice_action} with $c_1=1$ and $c_2=0$. 
The topological charge was evaluated using the geometric definition after 
a fixed number of $20$ cooling steps,
as its value was systematically observed to stabilize after $10$ steps. In the end, the topological susceptibility will then
be defined on a finite lattice as follows,
\beq
a^2 \chi &=& \frac{\braket{Q^2}}{L^2},
\eeq
where $Q \equiv Q_{\rm geo}^{(\rm cool)} \in \mathbb{Z}$ is the integer-valued lattice geometric topological charge computed after cooling, and where the meaning of the mean over the ensemble in our out-of-equilibrium setup will be clarified in the next section.

\subsection{Out-of-equilibrium evolutions and Jarzynski's equality}\label{sec:off_eq_evo}

In this subsection, we explain in detail how Jarzynski's equality enables us
to compute vacuum expectation values in the system with PBCs (the \emph{target}
distribution), starting from a sampling of the system with OBCs (the \emph{prior} distribution). 

Consider the following family of partition functions,
\beq\label{eq:family_distr}
\mathcal{Z}_{c(n)} \equiv \int [\dd \overline{z} \dd z \dd U] e^{-S_L^{(n)}[\overline{z},z,U]}~.
\eeq
Each one corresponds to a system described by action in Eq.~\eqref{eq:lattice_action} with boundary conditions specified 
by the parameter $c$. The prior system with OBCs will then correspond 
to $c(n=0)=0$, with partition function $\mathcal{Z}_0$, while the target 
system with PBCs will instead correspond to $c(n=\nstep)=1$, with
partition function $\mathcal{Z}_1 \equiv \mathcal{Z}$.

A sampling of the target distribution is obtained from a sampling of the 
prior distribution through out-of-equilibrium evolutions. Along the evolution,
the boundary condition parameter is gradually changed from $c=0$ to $c=1$.
The change is effected in $\nstep$ steps. At each step $n$, the change from 
$c(n)$ to $c(n+1)$ is followed by a number of
configuration updates\footnote{The update algorithm must satisfy detailed balance.}.
These combined operations allow us to define 
a transition probability distribution
$\mathcal{P}_{n}(\phi_{n-1} \to \phi_n)$ 
where $\phi_{n-1}$ and $\phi_n$ collectively denote the elementary degrees 
of freedom 
$z$ and $U$ at step $n-1$ and $n$, respectively.
As just a few updates are performed after every change in $c$, 
the system with elementary degrees of freedom $\phi_n$ can be 
considered to be out of equilibrium.

Jarzynski's equality allows to compute the ratio between the partition functions of the target and prior distributions,
\beq\label{eq:jarz_ratio_Z}
\frac{\mathcal{Z}}{\mathcal{Z}_0} = \braket{\exp\{-W\}}_{\mathrm{f}}~,
\eeq
where
\beq\label{eq:work}
 W[\phi_0, \dots, \phi_{\nstep-1}] \equiv \sum_{n=0}^{\nstep-1} 
 \left\{ S_L^{(n+1)}\left[\phi_n\right] - S_L^{(n)}\left[\phi_n\right] \right\}
\eeq
is known as \emph{generalized work}. The out-of-equilibrium
average $\langle \mathcal{A} \rangle_{\mathrm{f}}$ is defined as follows,
\beq\label{eq:evolution_average}
\langle \mathcal{A} \rangle_{\mathrm{f}} = \int [\dd \phi_0 \dots \dd \phi ] q_0[\phi_0] \, \mathcal{P}_{\mathrm{f}}[\phi_0,\dots, \phi] \, \mathcal{A}[\phi_0, \dots, \phi]~,
\eeq
where $\phi \equiv \phi_{\nstep}$ 
denotes the configuration reached at 
the end of the out-of-equilibrium evolution.
In the above, $\mathcal{P}_{\mathrm{f}}[\phi_0,\dots,\phi] \equiv \prod_{n=1}^{\nstep} 
\mathcal{P}_n(\phi_{n-1} \to \phi_n)$ is the total transition probability 
of the out-of-equilibrium evolution, and 
$q_0[\phi_0] \equiv \mathcal{Z}_0^{-1} \exp\{-S_L^{(0)}[\phi_0]\}$ 
the probability of drawing the field configuration $\phi_0$ from the prior distribution.

Using Eq.~\eqref{eq:jarz_ratio_Z} and Eq.~\eqref{eq:evolution_average},
one can show that the expectation value of an observable $\mathcal{O}$ 
with respect to the target distribution can be expressed as follows,
\beq\label{eq:rew_like_jarz}
\langle \mathcal{O} \rangle_{\NE} = \frac{\left\langle \mathcal{O}[\phi] \exp\{-W[\phi_0, \dots, \phi_{\nstep-1}]\} \right\rangle_{\mathrm{f}} }{\langle \exp\{-W[\phi_0, \dots, \phi_{\nstep-1}] \} \rangle_{\mathrm{f}} }
\eeq
where the ${\rm NE}$ denotes the fact that a Non-Equilibrium method has been used 
for the computation. 

In the following, our strategy will thus be to start from
an ensemble of configurations $\{ \phi_0 \}$, sampled from 
the prior distribution $q_0$, i.e., using 
action Eq.~\eqref{eq:lattice_action} with $c(0)=0$, 
and to perform, for each configuration, the out-of-equilibrium 
evolution defined above. Of course, there is no unique way to interpolate
from $c=0$ to $c=1$. Hence, we supplement the above procedure with
a \emph{protocol},
\beq\label{eq:protocol}
\phi_0 \underset{c(0)\,\to\,c(1)}{\longrightarrow} 
\phi_1 \underset{c(1)\,\to\,c(2)}{\longrightarrow} \phi_2 \quad \dots \quad
\underset{c(\nstep-1)\,\to\,c(\nstep)}{\phi_{\nstep-1} \quad \longrightarrow \quad \phi_{\nstep}\equiv \phi}~,
\eeq
that is, with a choice for the value of $c(n)$ for each step.

\begin{figure}[!t]
\centering
\includegraphics[scale=0.7,keepaspectratio=true]{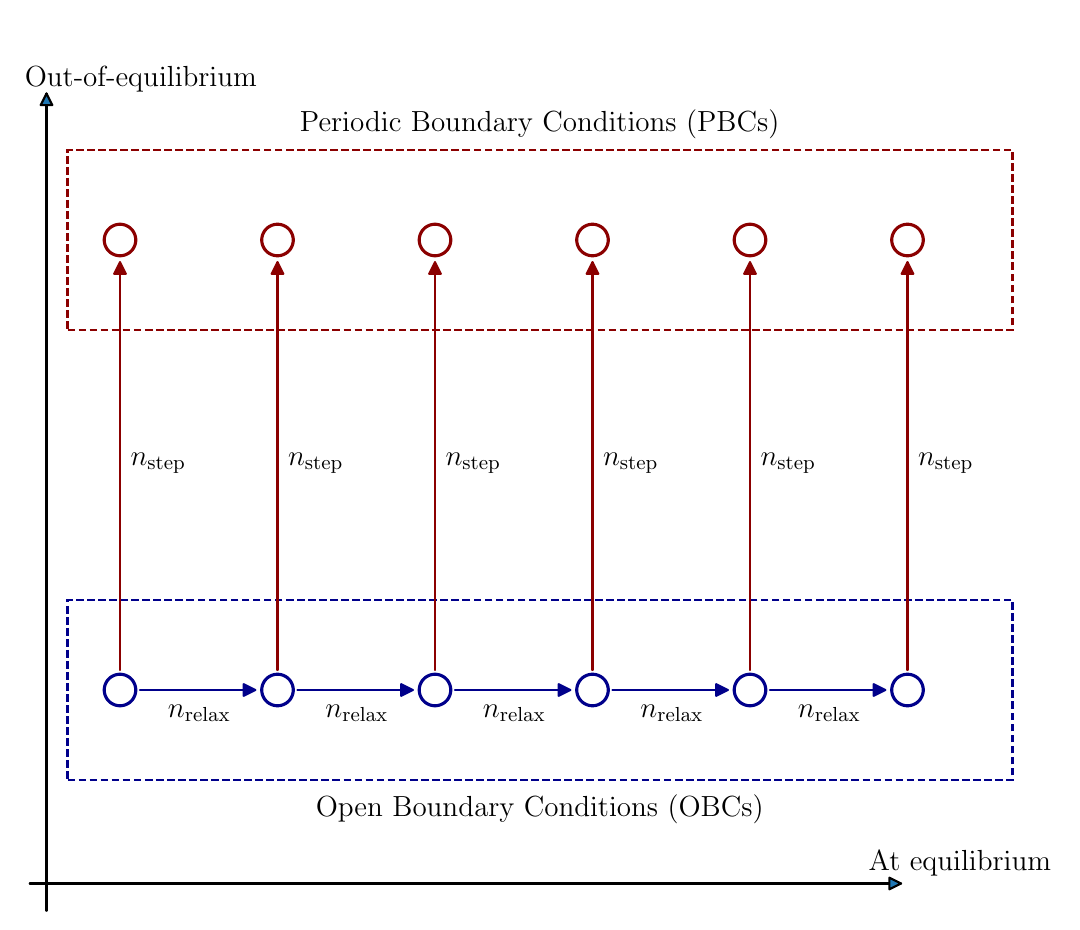}
\caption{Sketch of the out-of-equilibrium setup. 
The horizontal axis represents the Monte Carlo time, where a new configuration is 
generated at equilibrium every $\nrelax$ updating steps according to the 
prior distribution, i.e., with OBCs. 
The vertical arrows instead stand for the $\nev$ out-of-equilibrium 
evolutions used to gradually switch on PBCs, each of these $\nstep$ steps long. 
The bullets correspond to the states of the system. 
Observables are computed only at the end of the out-of-equilibrium 
evolution, while the work is computed all along.}
\label{fig:off_evo}
\end{figure}

While it is necessary to perform a measurement of the action at each step 
in order to obtain the generalized work from Eq.~\eqref{eq:work}, the
measurement of the observable of interest $\mathcal{O}$ is only
performed at the end of the out-of-equilibrium evolution. In other
words, the observable is only measured when the system has reached PBCs, thus avoiding unphysical effects introduced by OBCs. A sketch of our overall algorithmic procedure is displayed in Fig.~\ref{fig:off_evo}.

As is also evident from Eq.~\eqref{eq:rew_like_jarz}, the technique laid out 
above has conceptual similarities with common reweighting techniques. It is 
natural to expect the \emph{quality} of sampling of the target 
distribution to depend on the overlap between the successive distributions 
in $\mathcal{Z}_c$, which is expected to be smaller for
evolutions that are farther from equilibrium. Since too far 
out-of-equilibrium evolutions could lead to a small signal-to-noise ratio 
and/or to a possibly large bias in the computation of 
$\braket{\mathcal{O}}_{\NE}$ through Eq.~\eqref{eq:rew_like_jarz}, 
it is important to quantify the distance of the chosen evolution from equilibrium.

A figure of merit designed to fit that purpose is the reverse
Kullback--Leibler divergence $\DKL$, which 
is a measure of the similarity between two probability 
distributions and it is defined to be always 
larger or equal to zero. 
In general, the simplest choice would be to compute the overlap 
between the target distribution and the one 
that has been generated at the end of an out-of-equilibrium evolution. 
The corresponding Kullback--Leibler divergence is:
\beq
\label{eq:KL_div_0}
    \DKL (q \| p) = \int \dd \phi \; q (\phi) \log \left( \frac{q (\phi)}{p (\phi) } \right),
\eeq
where $p\equiv \mathcal{Z}^{-1} \exp\{ S_L^{(\nstep)} \}$ is the probability of drawing a configuration from the target distribution, while $q$ is a (generally intractable) distribution of the form
\beq
    q(\phi) = \int [\dd \phi_0 \dots \dd \phi_{\nstep -1 } ] q_0[\phi_0] \mathcal{P}_{\mathrm{f}}[\phi_0,\dots,\phi].
\eeq

For out-of-equilibrium evolutions we have no direct access to $q$ and this
prevents us from computing the quantity in Eq.~\eqref{eq:KL_div_0}. However, 
another, different, Kullback--Leibler divergence can be defined
by comparing the forward and reverse transition probabilities between 
$\phi_0$ and $\phi$.  Labeling the former by $\mathrm{f}$, 
see Eq.~\eqref{eq:protocol}, and the latter by $\mathrm{r}$, we have:
\beq\label{eq:KL_div}
\DKL(q_0 \mathcal{P}_{\mathrm{f}} \| p \mathcal{P}_{\mathrm{r}}) &=& \int [\dd \phi_0 \dots \dd \phi ] q_0[\phi_0] \mathcal{P}_{\mathrm{f}}[\phi_0,\dots,\phi] \log \left( \frac{q_0[\phi_0] \mathcal{P}_{\mathrm{f}}[\phi_0,\dots,\phi]}{p[\phi] \mathcal{P}_{\mathrm{r}}[\phi,\dots,\phi_0]} \right).
\eeq
This quantity is directly related to thermodynamic quantities: if $\Delta F \equiv -\log \frac{\mathcal{Z}}{\mathcal{Z}_0}$ is the variation of the free energy along the out-of-equilibrium evolution and $W$ the corresponding (generalized) work from Eq.~\eqref{eq:work}, then it is easy to derive (using detailed balance) that
\beq\label{eq:DKL_therm_meaning}
\DKL(q_0 \mathcal{P}_{\mathrm{f}} \| p \mathcal{P}_{\mathrm{r}}) = \langle W \rangle_{\mbox{\tiny{f}}} + \log \frac{\mathcal{Z}}{\mathcal{Z}_0} = \langle W \rangle_{\mbox{\tiny{f}}} - \Delta F~ \geq 0.
\eeq
For an equilibrium process, for which $\DKL=0$, we have that $\langle W \rangle_{\mbox{\tiny{f}}} = \Delta F$, i.e., all of the work spent in the out-of-equilibrium evolution is transferred into the free energy difference between the prior and the target distributions;
this is exactly the expectation for a reversible evolution through equilibrium states, for which there is no difference between forward and reverse.
For a generic out-of-equilibrium process we have instead $\DKL > 0$, that corresponds to evolutions for which the probabilities of the forward and reverse evolutions are different. 
From a thermodynamic perspective this implies, instead, $\langle W \rangle_{\mbox{\tiny{f}}} > \Delta F$, i.e., part of the work is dissipated. It is then clear that Eq.~\eqref{eq:DKL_therm_meaning} is a restatement of the Second Principle of Thermodynamics.
Finally, it is easy to prove that:
\beq
\DKL (q \| p) \leq \DKL(q_0 \mathcal{P}_{\mathrm{f}} \| p \mathcal{P}_{\mathrm{r}}).
\eeq
Thus, the value of the divergence of Eq.~\eqref{eq:KL_div} also puts a
constraint on how far the actual generated distribution $q$ is from the 
target distribution $p$.

Another figure of merit that can be used to quantify the distance from 
equilibrium is the so-called \emph{Effective Sample Size} (ESS). 
This quantity is customarily employed in the context of 
(Stochastic) Normalizing Flows as it encapsulates the relationship between 
the variance of an observable sampled 
directly from the target distribution $p$ and the variance of the same 
observable obtained using Eq.~\eqref{eq:rew_like_jarz}. Ignoring auto-correlations, 
it is easy to show that
\beq
\label{eq:ESS_var}
\frac{\Var(\mathcal{O})_{\NE}}{n} = \frac{\Var(\mathcal{O})_p }{n \, \ESS}~,
\eeq
which also motivates the name \emph{Effective Sample Size}. The variance of 
an observable $\mathcal{O}$, obtained through Eq.~\eqref{eq:rew_like_jarz}, 
is equal to the variance obtained from the target distribution
$p$ with a smaller sample of size $n_{\mathrm{eff}} = \ESS \times n\leq n$.

In this study, we employ the following estimator:
\beq
\label{eq:ESS_def}
\hat \ESS \equiv \frac{\langle e^{-W} \rangle_{\mathrm{f}}^2}{ \langle e^{-2W} \rangle_{\mathrm{f}}} = \frac{1}{\langle e^{-2(W-\Delta F)} \rangle_{\mathrm{f}}}~,
\eeq
and we refer to Ref.~\cite{elvira2022rethinking} for a discussion on how $\hat \ESS$ is related to the true Effective Sample Size of Eq.~\eqref{eq:ESS_var}.
The estimator $\hat\ESS$ can be easily related to the variance of the weights $e^{-W}$ appearing 
in Eq.~\eqref{eq:rew_like_jarz}. Indeed, since $\Var(e^{-W}) = \braket{e^{-2W}}_{\mathrm{f}} - \braket{e^{-W}}_{\mathrm{f}}^2 \geq 0$, then %
\beq
\label{eq:ESS_expwvar}
\Var(e^{-W}) = \left(\dfrac{1}{\hat\ESS}-1\right) \braket{e^{-W}}_{\mathrm{f}}^2 = \left(\dfrac{1}{\hat\ESS}-1\right) e^{-2\Delta F} \geq 0~,
\eeq
and, as a consequence,
\beq
0 < \hat \ESS \leq 1~.
\eeq
From the above we see that $\hat{\ESS}=1$ implies $\Var(e^{-W})=0$, which is only 
possible if the weights $e^{-W}$ are all equal. In this case, it is apparent
from Eq.~\eqref{eq:rew_like_jarz} that no reweighting is being done at all, 
and we are at equilibrium. On the other hand, a value of $\hat{\ESS}$ approaching
zero signals sizeable fluctuations in the weights $e^{-W}$, 
corresponding to a noisy reweighting and to out-of-equilibrium evolutions. 

\section{Numerical results}\label{sec:res}

This section is devoted to the discussion of the numerical results concerning
the efficiency of the use of out-of-equilibrium evolutions, which will 
be measured in terms of the auto-correlation time of $\chi$. 

In Ref.~\cite{Hasenbusch:2017unr}, the PTBC algorithm was shown to outperform standard
local algorithms both in the presence of PBCs and OBCs. Moreover, it was shown to 
enjoy smaller auto-correlation times with respect to other setups, such as 
metadynamics simulations, see Ref.~\cite{Laio:2015era}. Hence, in the following, 
a direct comparison between the performance of the out-of-equilibrium protocol and of
parallel tempering will be performed.

Since our goal is to test the robustness of the method, we have chosen 
to probe a wide array of different combinations of 
$L_d$, $\nstep$ and $\nrelax$ in independent simulations, rather
than obtaining larger samples for a smaller number of setups.

Thus, simulations for two different values of the parameter $N$
specifying the model were performed, and we focused, for each value of these,
on a single value of $\beta$ and $L$, 
see Tab.~\ref{tab:setup}. These
simulation setups were chosen to enable direct comparison with
previous results from Refs.~\cite{Berni:2019bch,Bonanno:2022yjr}. However,
for a few choices of $L_d$ and $\nstep$ we will also present results for the
auto-correlation time obtained with our out-of-equilibrium setup varying the
lattice spacing and/or the volume.

\begin{table}
\centering
\begin{tabular}{|cccccc|}
\hline
$N$ & $\beta$ & $L$ & $L_d$ & $\nstep$ & $\nrelax$ \\
\hline
21 & 0.7 & 114 & [6,114] & [200,2000] & [50,250]\\
41 & 0.65 & 132 & [10,30] & [500,2000] & [50,250]\\
\hline
\end{tabular}
\caption{Setup of the numerical simulations. The model is specified by $N$, $\beta$ and the size of the square lattice $L$. Non-equilibrium evolutions are characterized by the choice of the length $L_d$ of the line defect, the length of the evolution $\nstep$ and the intermediate number of updating steps between each out-of-equilibrium evolution $\nrelax$.}
\label{tab:setup}
\end{table}

\subsection{Effective Sample Size and Kullback--Leibler divergence}

As a first step in the investigation of the performance of our new proposal, we
study how both the Kullback--Leibler divergence $\DKL$ and the Effective Sample
Size $\hat{\ESS}$ depend on $\nstep$ and on the defect length
$L_d$.

The aim of this section is, in particular, to quantify the distance of Jarzynski
evolutions from equilibrium as the parameters $\nstep$ and $L_d$ are varied.
This is a crucial step in order to assess the reliability of the exponential
Jarzynski reweighting.

\begin{figure}[!t]
\centering
\includegraphics[scale=0.45,keepaspectratio=true]{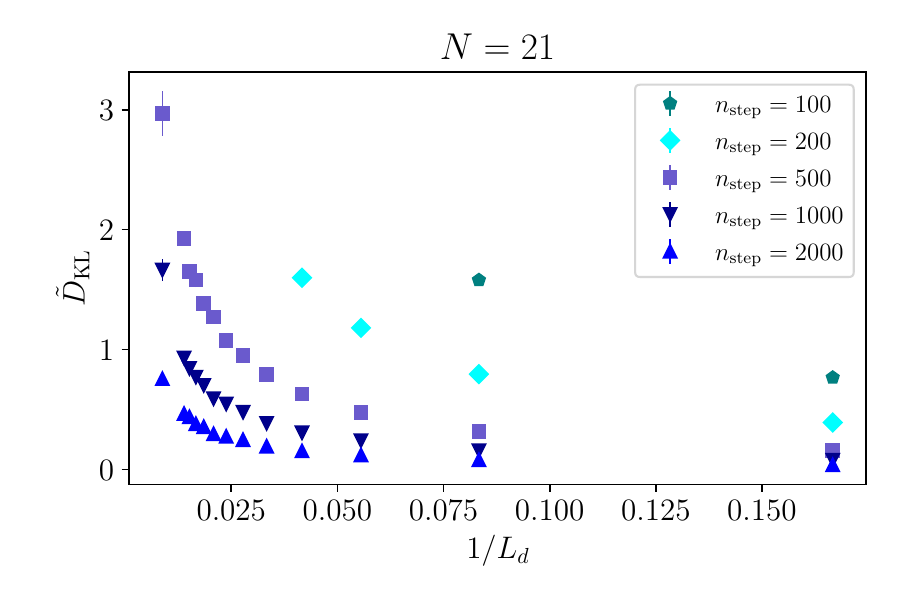}
\includegraphics[scale=0.45,keepaspectratio=true]{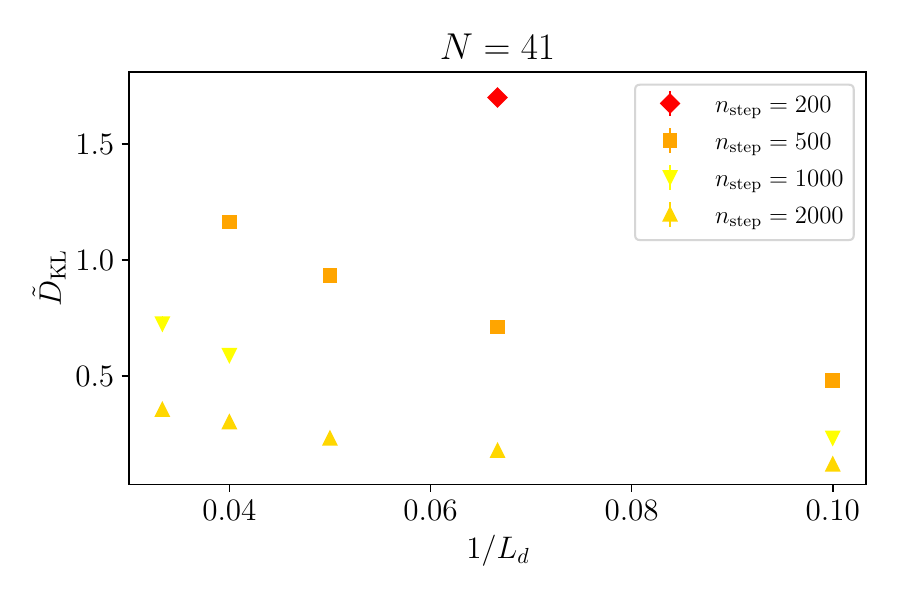}
\\
\includegraphics[scale=0.45,keepaspectratio=true]{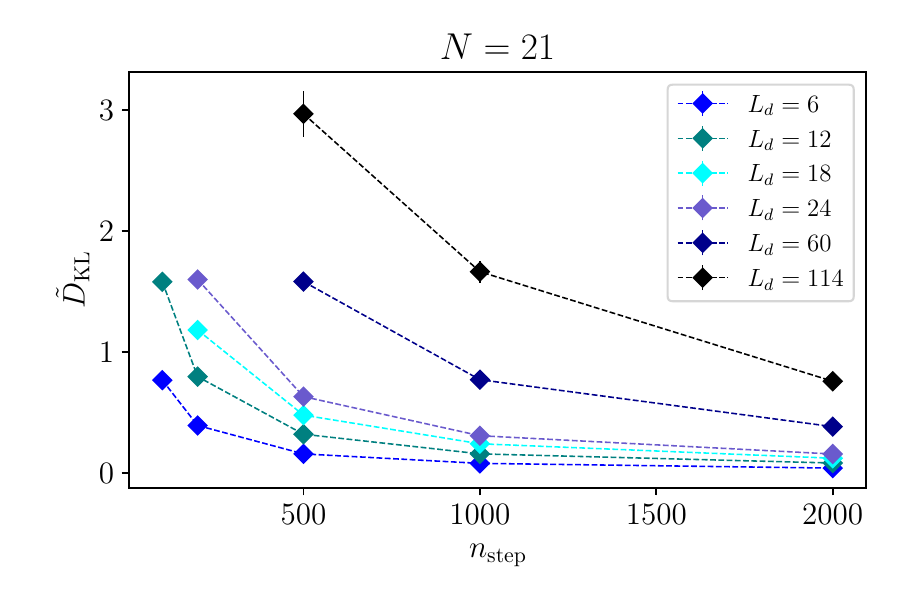}
\includegraphics[scale=0.45,keepaspectratio=true]{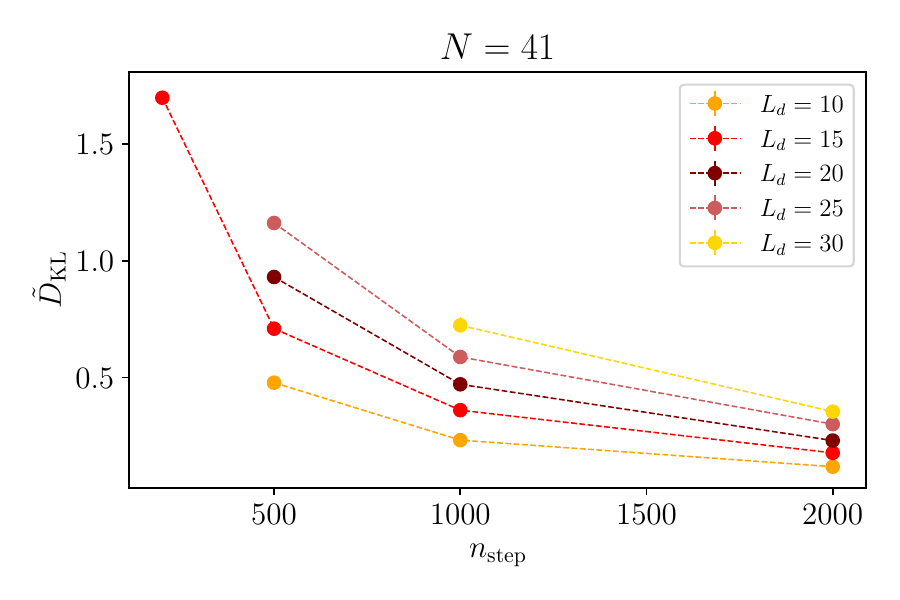}
\caption{Behavior of the $\DKL$ as a function of $\nstep$ (lower panels) and of the inverse of the defect size $L_d$ (upper panels), for $N=21$ (left panels) and $N=41$ (right panels).}
\label{fig:DKL}
\end{figure}

On general grounds, we expect the Kullback--Leibler divergence to approach
zero, its expected value at equilibrium, when either 
$\nstep$ is increased at fixed $L_d$ or $L_d$ is decreased at fixed $\nstep$. 
The expectation explained above is clearly 
confirmed in Fig.~\ref{fig:DKL}, where $\DKL$ is displayed as a 
function of $1/L_d$ and $\nstep$ for fixed $\nrelax$.

\begin{figure}[!t]
\centering
\includegraphics[scale=0.45,keepaspectratio=true]{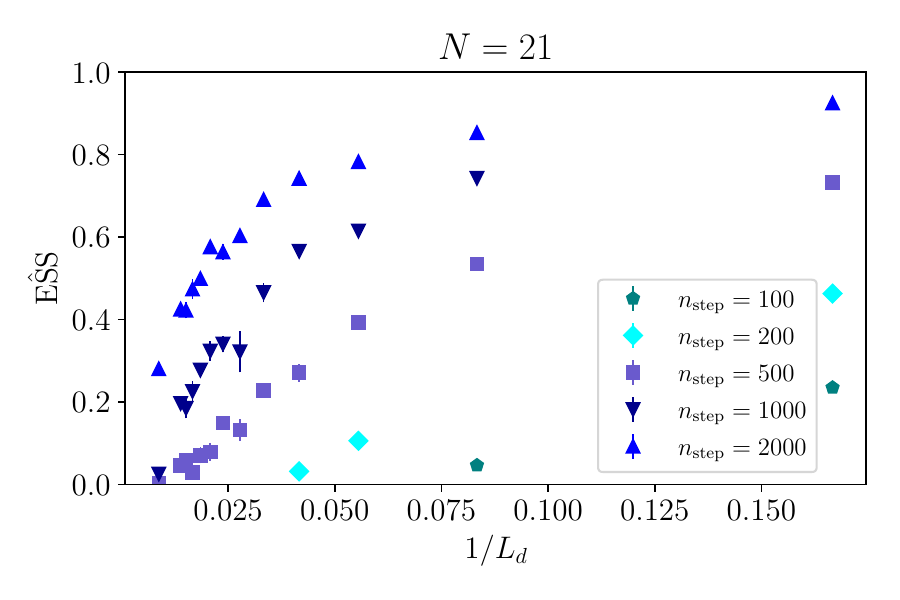}
\includegraphics[scale=0.45,keepaspectratio=true]{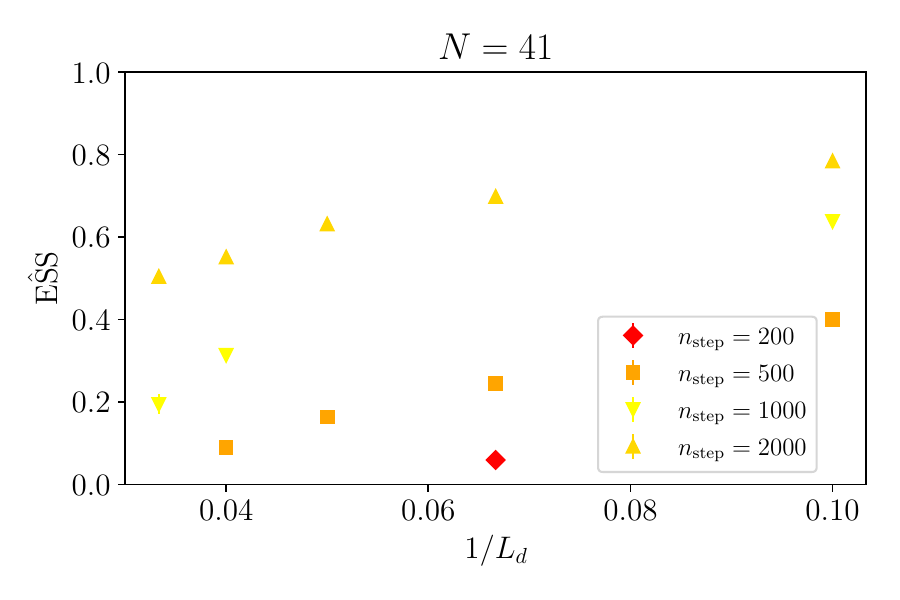}
\\
\includegraphics[scale=0.45,keepaspectratio=true]{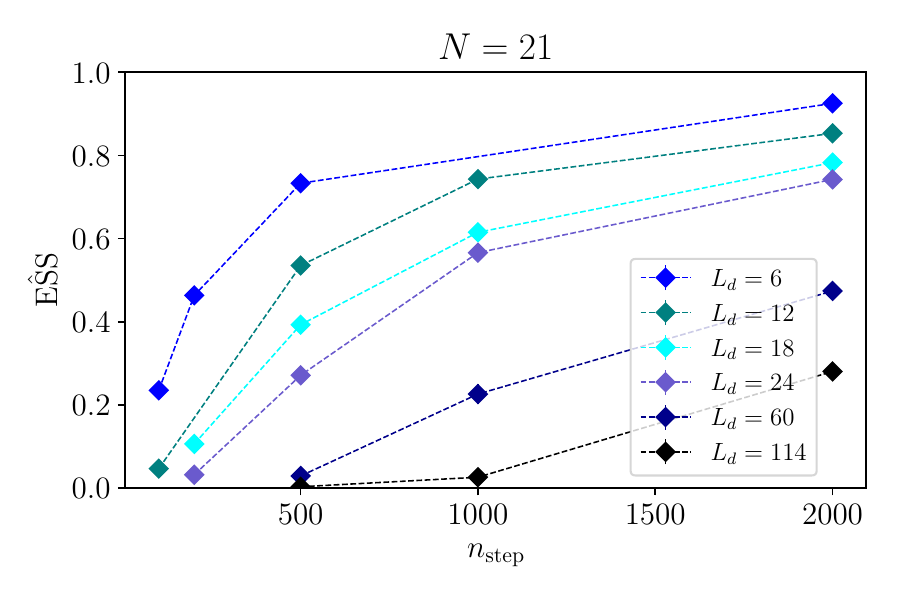}
\includegraphics[scale=0.45,keepaspectratio=true]{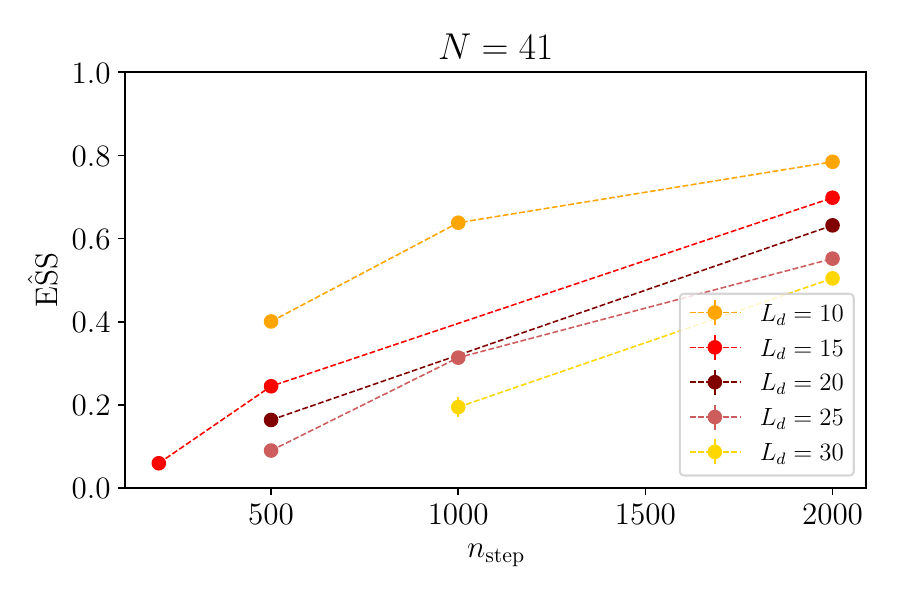}
\caption{Behavior of the $\hat{\ESS}$ as a function of $\nstep$ (lower panels) and of the inverse of the defect size $L_d$ (upper panels), for $N=21$ (left panels) and $N=41$ (right panels).}
\label{fig:ESS}
\end{figure}

The behaviour of $\hat{\ESS}$ reflects the same picture.
The evolution towards the target distribution is expected to approach
equilibrium as $\nstep$ is increased and to recede from it
as $L_d$ is increased at fixed $\nstep$. Accordingly,
$\hat{\ESS}$ is expected to approach $1$ in the former case, and to recede
from it in the latter. This is indeed what can be observed in Fig.~\ref{fig:ESS},
where $\hat{\ESS}$ is displayed as a function of $1/L_d$ and of $\nstep$, in the
left and right panels, respectively, for fixed $\nrelax$.

For sufficiently small defects, $L_d<20$, the value of $\hat{\ESS}$ is 
seen to become greater than $0.5$ already for $\nstep \simeq 500$. 
For larger defects, $L_d>20$, a value of $\nstep$ of order 1000--2000
is required for $\hat{\ESS}$ to be greater than $0.5$.
While $\hat{\ESS}=0.5$ can be considered as a lower safety threshold,
we will see below that the Monte Carlo ensembles that were employed were large
enough to allow an acceptable computation of the topological
susceptibility with even smaller values.

\begin{figure}[!t]
\centering
\includegraphics[scale=0.45,keepaspectratio=true]{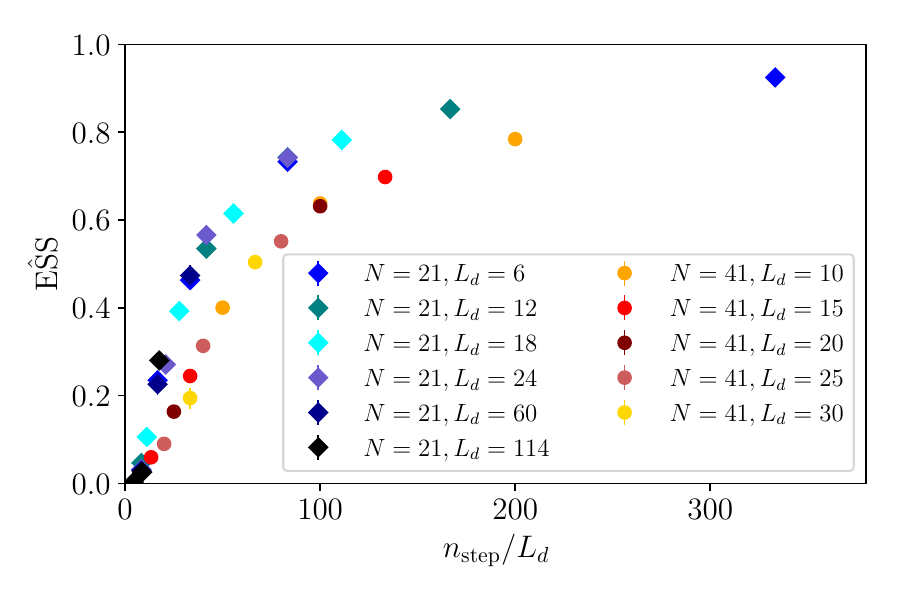}
\includegraphics[scale=0.45,keepaspectratio=true]{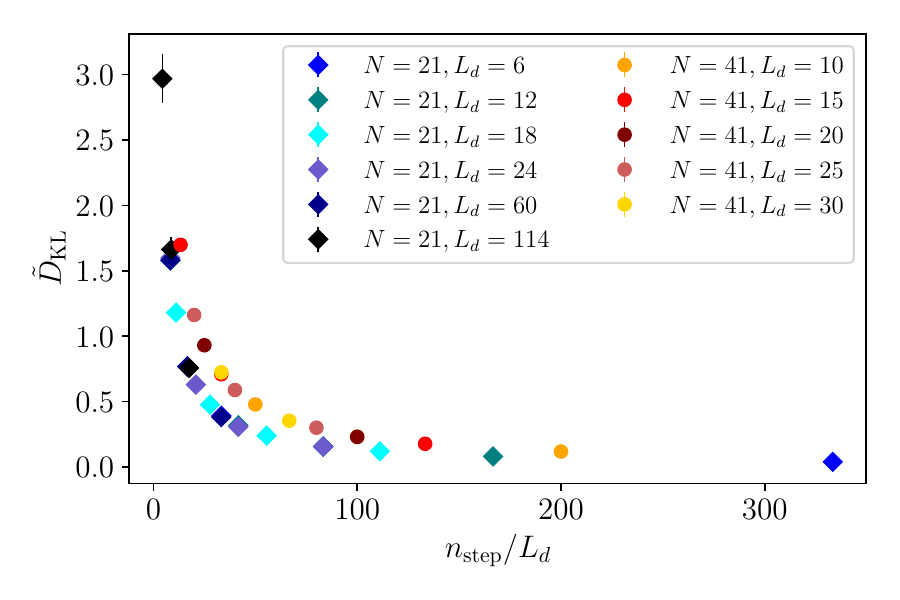}
\caption{Behavior of the $\hat{\ESS}$ (left panel) and of $\DKL$ (right panel) as a function of $\nstep/L_d$ at fixed value of $\nrelax$.}
\label{fig:ESSDKL_nstepd}
\end{figure}

A careful inspection of the available data suggest that, in fact,
both $\hat{\ESS}$ and $\DKL$ are, to a good approximation,
only functions of the ratio $\nstep/L_d$. That this is a sensible
idea can be immediately appreciated from Fig.~\ref{fig:ESSDKL_nstepd},
where $\hat{\ESS}$ and $\DKL$ are shown to collapse on two different
single $\nstep/L_d$ dependent curves at two different values of $N$.
A semi-quantitative justification can instead be obtained
from the definition of work in Eq.~\eqref{eq:work}. 
Indeed, the calculation of $W$ involves
the sum of the variations of the action induced by 
the change $c(n)\longrightarrow c(n+1)$.
Focusing for simplicity on only the nearest-neighbors 
interaction terms, we can write:
\beq\label{eq:argument_nstepd}
S_L^{(n+1)}[\phi_n]-S_L^{(n)}[\phi_n] & \propto &\sum_{\underset{\underset{\mu=0}{x_0=L-1}}{x_1=0}}^{L_d-1} \Delta c(n)\Re\left[U_\mu(x)\overline{z}(x+\hat{\mu})z(x)\right],
\eeq
with
\beq
\Delta c(n) &=& c(n+1)-c(n) = \frac{1}{\nstep},
\eeq
and where we have assumed that the defect lies on the $x_0=L-1$ boundary from
$x_1=0$ to $x_1=L_d-1$. 
Since a difference of actions is an extensive quantity and is,
in the case at hand, localized on the defect, then the average
of the quantity in Eq.~\eqref{eq:argument_nstepd} is of order 
$L_d$. Then, as a consequence of our choice of interpolating function, 
$c(n)=n/\nstep$, it is clear that Eq.~\eqref{eq:argument_nstepd} 
is, on average, only a function of $L_d\times \Delta c(n) = L_d/\nstep$. 
The same considerations can be made on the
next-to-nearest-neighbors interaction term. Hence, the work $W$ will
on average only depend on $\nstep/L_d$ and it is natural to 
think that the same holds true for $\hat{\ESS}$ and $\DKL$ as well.

\begin{figure}[!t]
\centering
\includegraphics[scale=0.45,keepaspectratio=true]{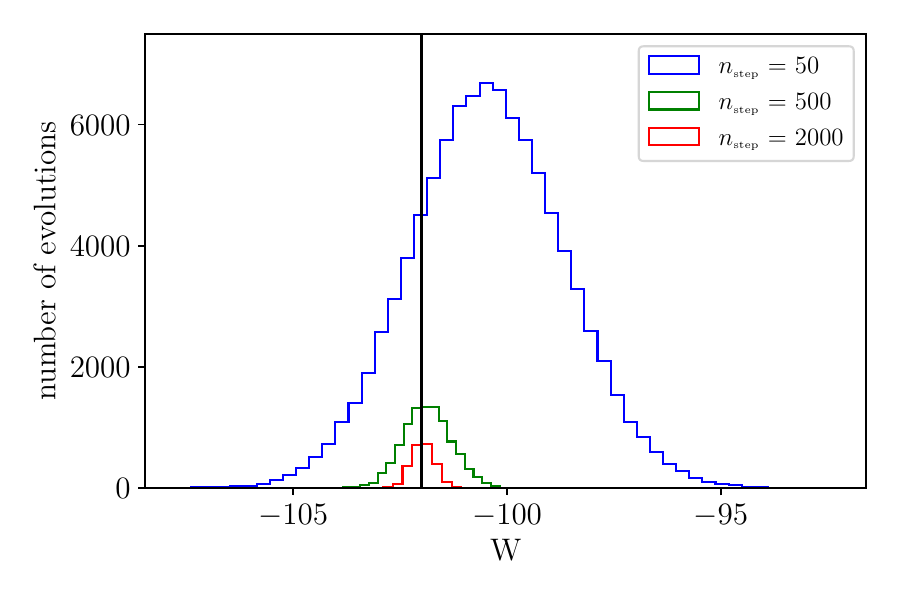}
\includegraphics[scale=0.45,keepaspectratio=true]{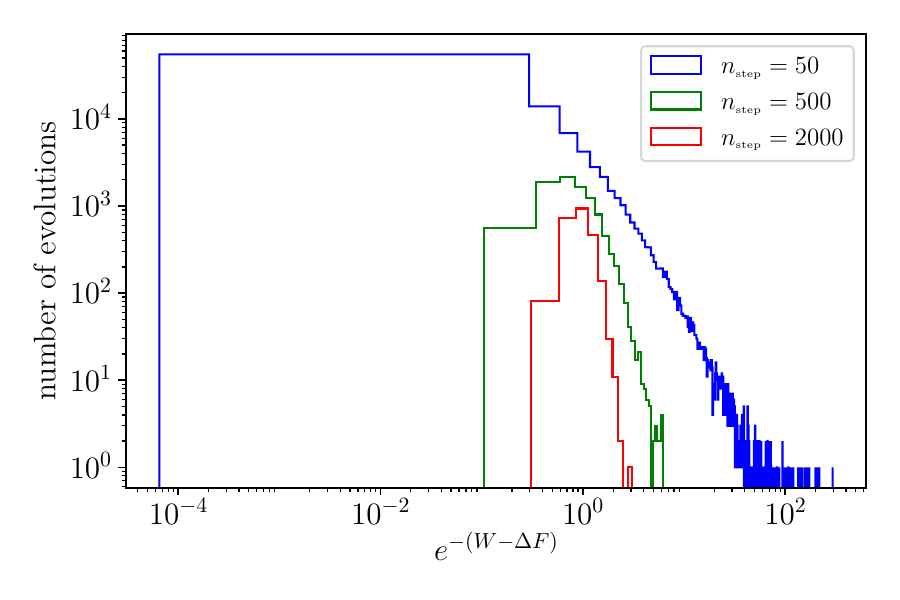}
\caption{Distribution of the work $W$ (left panel) and of $\exp\{-(W-\Delta F)\}$ (right panel) for $L_d=6$ for various values of $\nstep$. In the left panel the vertical bar represents the value of $\Delta F$ for these particular evolutions.}
\label{fig:histograms}
\end{figure}

Further insight on $\hat{\ESS}$ and $\DKL$ can be obtained from
Fig.~\ref{fig:histograms} where the frequency histograms of the 
out-of-equilibrium evolution are displayed as functions of $W$ and 
of the normalized\footnote{The weight $w$ is normalized so that $\braket{w}_{\rm f}  = 1$ by virtue of Jarnzyski's equality.} 
weight 
$w=\exp\{-W + \Delta F\}$
that appears in Eq.~\eqref{eq:rew_like_jarz}.

Three different cases are represented, corresponding to evolutions
that are far from ($\nstep=50$), \emph{moderately} far from ($\nstep=500$) and
close
($\nstep=2000$) to equilibrium.
The left-hand panel of the figure shows how
the peak in the distribution of the evolutions in $W$ 
approaches the value of $\Delta F$, represented as the vertical black line,
as $\nstep$ grows, i.e. as the evolution approaches equilibrium. 
This corresponds, in the right-hand panel, to the distribution
of evolutions in $w$ becoming progressively more peaked around $1$,
where $W=\Delta F$, with the variance in $w$ approaching zero.

The right-hand panel also enables us to better understand
the reliability of the statistical reweighting procedure in 
Eq.~\eqref{eq:rew_like_jarz}. When the protocol is relatively far
from equilibrium ($\nstep=50$), the distribution in $W$ has a larger 
support and, correspondigly, the support of the distribution in $w$ 
then extends exponentially to smaller values. 
In this case, large statistics would be necessary to provide an 
unbiased sampling of $w$.
To the opposite, a more peaked distribution in $W$
is obtained as $\nstep$ is incrased, corresponding to a smaller support 
in the distribution of $w$, and to an easier sampling. This is completely 
analogous to how the reliability of the re-weighting technique in classical
statistical mechanics depends on the overlap between the supports of the
source and target distributions.

In conclusion, both the $\hat{\ESS}$ and $\DKL$ behave as expected on
theoretical grounds and according to the general discussion in
Sec.~\ref{sec:off_eq_evo}. This means, in practice, that we are able to control
the magnitude of systematic effects originated from the reweighting step
in Eq.~\eqref{eq:rew_like_jarz}, and we can fully trust
the expectation values obtained for the target distribution. In the 
next section, we will present a practical application of the strategy
laid out above. Further details on the mutual relation between
$\hat{\ESS}$ and $\DKL$ can be found in Appendix~\ref{app:DKL_vs_ESS}.

\FloatBarrier

\subsection{Extracting the topological susceptibility}

\begin{figure}[!t]
\centering
\includegraphics[scale=0.48,keepaspectratio=true]{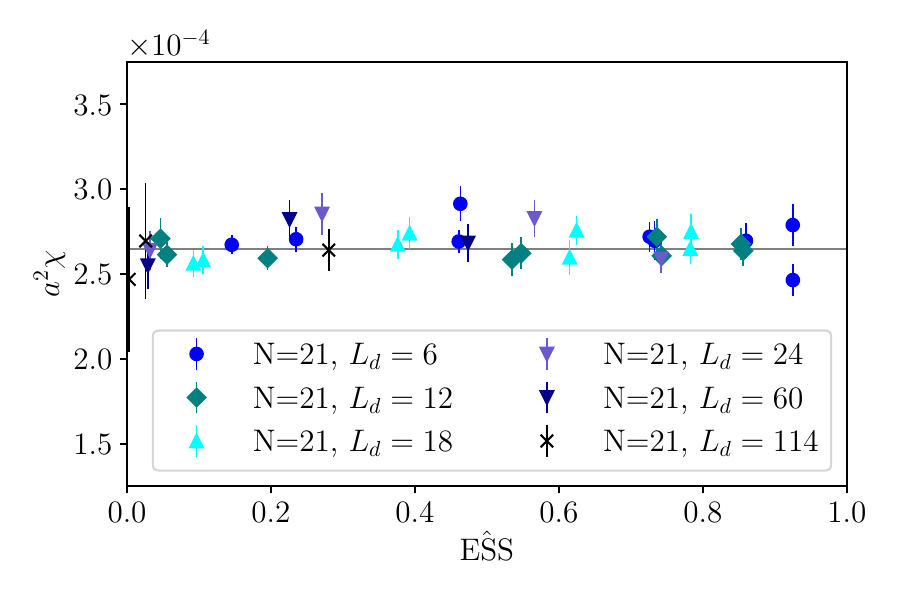}
\includegraphics[scale=0.48,keepaspectratio=true]{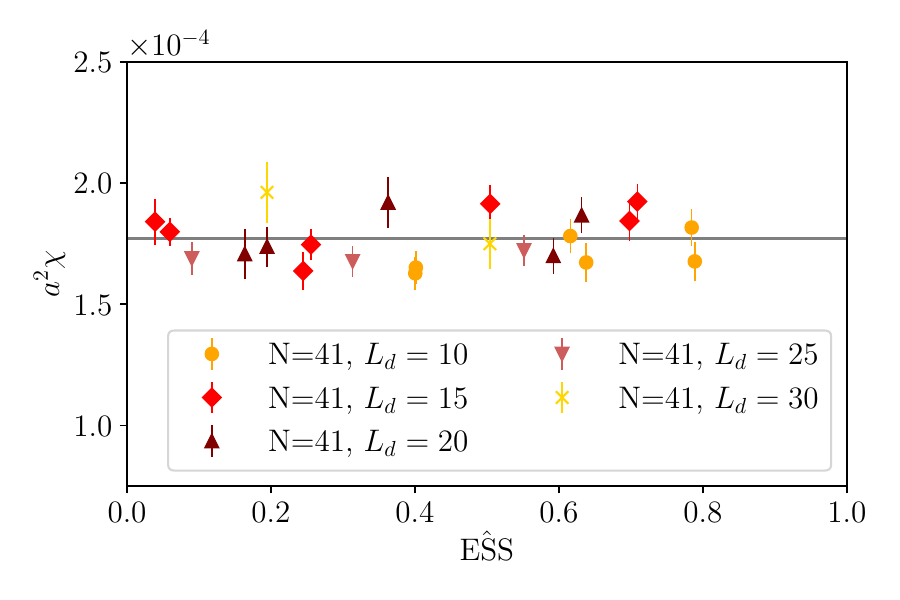}
\caption{Results for the topological susceptibility in lattice units $a^2 \chi$ obtained with non-equilibrium evolutions as a function of the Effective Sample Size, for $N=21$ and $\beta=0.7$ (left panel) and for $N=41$ and $\beta=0.65$ (right panel). The results obtained with the PTBC algorithm in Ref.~\cite{Berni:2019bch} is reported as a black horizontal line.}
\label{fig:chiESS}
\end{figure}

A necessary condition for the viability of the non-equilibrium methods
must of course rely on the comparison between its results and 
the results obtained with equilibrium methods. 
In Fig.~\ref{fig:chiESS} the values of the topological susceptibility
$\chi$ obtained using Eq.~\eqref{eq:rew_like_jarz} and several
combinations of $L_d$, $\nstep$ and $N$ are displayed,
along with the values obtained in Ref.~\cite{Berni:2019bch}, using the PTBC
algorithm.

The non-equilibrium results are found to be in excellent agreement with
those from the PTBC algorithm for the whole range of values of $\hat{\ESS}$. 
Notably, no bias seems to arise even when $\hat{\ESS}\sim 0.1$ or smaller,
while for nearly vanishing $\hat{\ESS}$, an increase in
the magnitude of the error seems to occur for $L_d=114$ at $N=21$.
This remarkable fact can be understood in terms of the computational
effort spent to obtain each estimate of $a^2\chi$. The computational
effort is of course proportional to $\nstat =
\nev\times(\nstep+\nrelax)$, and this is roughly the same
for each point on the plot in Fig~\ref{fig:chiESS}. Hence, the data
on the lower range of $\hat{\ESS}$, for which $\nstep/L_d$ is small,
is characterized by a correspondingly larger factor $\nev$.
This is in agreement with the previous observation that a comparatively 
larger statistics is needed to avoid bias in expectation values computed
from Eq.~\ref{eq:rew_like_jarz}, when using evolutions that 
are farther from equilibrium.

\subsection{Integrated auto-correlation time of $\chi$}

A quantitative study of the non-equilibrium method crucially
revolves around the computation of the integrated auto-correlation time
of the topological susceptibility. The auto-correlation function for $\chi$,
from which the integrated auto-correlation time will be calculated, 
has been computed as follows,
\beq
\Gamma(t) \equiv \braket{ (Q^2_{i+t}/L^2 - a^2\chi)(Q^2_i/L^2 - a^2\chi)}_{\rm NE}
\eeq
where $Q^2_i$ denotes the cooled squared geometric lattice topological 
charge at the end  of the $i^{\text{th}}$
out-of-equilibrium evolution, i.e., when the system is subject to PBCs,
and
\beq
a^2 \chi = \frac{\braket{Q^2}_{\rm NE}}{L^2}.
\eeq
is the topological susceptibility computed from out-of-equilibrium evolutions.

\begin{figure}[!t]
\centering
\includegraphics[scale=0.48,keepaspectratio=true]{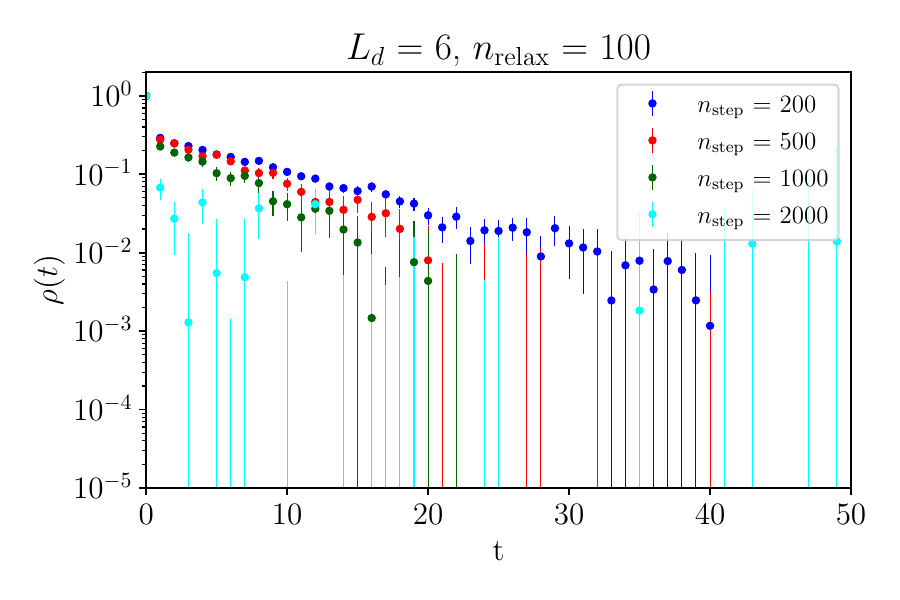}
\includegraphics[scale=0.48,keepaspectratio=true]{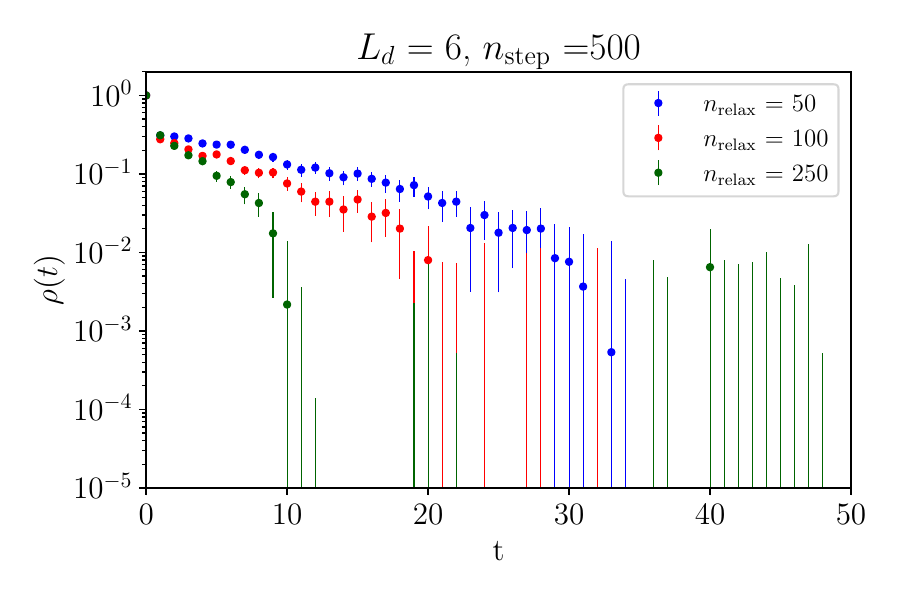}
\caption{Behavior of the normalized auto-correlation function $\rho(t)$ for $L_d=6$, for several values of $\nstep$ at fixed $\nrelax=100$ (left panel) and for several values of $\nrelax$ at fixed $\nstep=500$ (right panel).}
\label{fig:autocorr}
\end{figure}

For convenience, it is useful to define \emph{normalized} auto-correlation
function
$\rho(t) \equiv \Gamma(t) / \Gamma(0)$, which is displayed
in Fig.~\ref{fig:autocorr}
for various combinations of $\nstep$ and $\nrelax$ at fixed $L_d$. The integrated auto-correlation time is
defined in terms of $\rho(t)$ as follows,
\beq
\tauint = \frac{1}{2} + \sum_{t=1}^{\widetilde{W}} \rho(t),
\eeq
where $\widetilde{W}$ is the window computed using the $\Gamma$-method, see
Refs.~\cite{Wolff:2003sm,Joswig:2022qfe}. 

\begin{figure}[!t]
\centering
\includegraphics[scale=0.75,keepaspectratio=true]{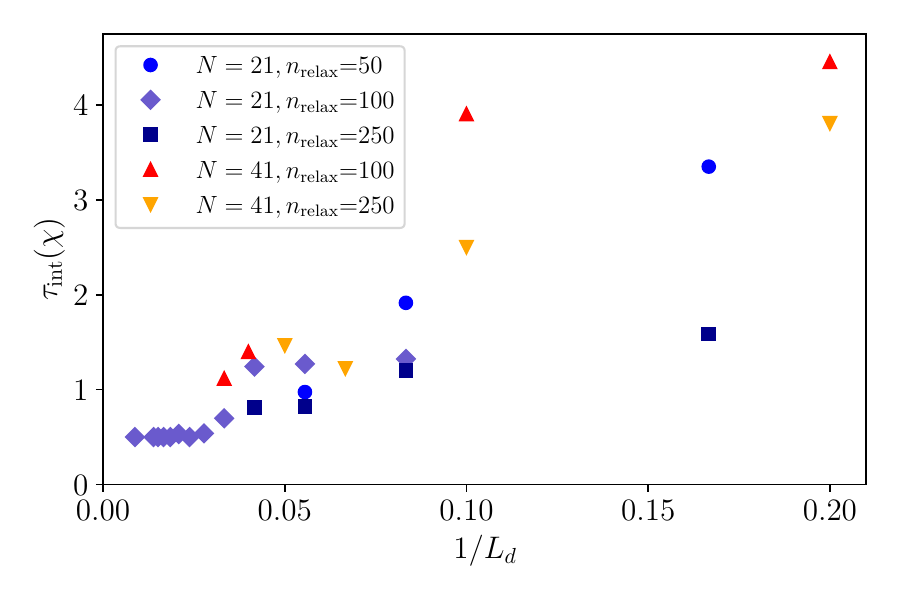}
\caption{Behavior of the integrated auto-correlation time $\tauint(\chi)$ as a function of $1/ L_d$ for $\nstep=1000$ and several values of $\nrelax$, for the $N=21$ and the $N=41$ ensembles.}
\label{fig:taud}
\end{figure}

The value of $\tauint$ computed for several 
combinations of $\nrelax$, $L_d$ are reported in Tab.~\ref{tab:tauint},
and displayed as a function of $1/L_d$, for $\nstep=1000$, in
Fig.~\ref{fig:taud}.
As one could expect, smaller values of $\tauint$ are generally 
observed for larger values of $\nrelax$, $L_d$ and $\nstep$, the latter
corresponding to evolutions closer to equilibrium.

\begin{table}
\centering
\begin{tabular}{|cccc|}
\hline
$L_d$ & $\nstep$ & $\nrelax$ & $\tauint$ \\
\hline
6 & 500 & 50 & 5.3(8)\\ 
6 & 500 & 100 & 3.6(5)\\
6 & 500 & 250 & 1.9(2)\\
6 & 1000 & 100 & 2.8(4)\\
12 & 1000 & 50 & 1.9(3)\\
12 & 1000 & 100 & 1.3(1)\\
12 & 1000 & 250 & 1.2(1)\\
24 & 1000 & 50 & 0.64(4)\\
24 & 1000 & 100 & 1.2(1)\\
24 & 1000 & 250 & 0.81(4)\\
36 & 1000 & 100 & 0.54(3)\\
\hline
\end{tabular}
\quad
\begin{tabular}{|cccc|}
\hline
$L_d$ & $\nstep$ & $\nrelax$ & $\tauint$ \\
\hline
10 & 500 & 100 & 3.4(4)\\
10 & 500 & 250 & 2.1(2)\\
10 & 1000 & 100 & 3.9(6)\\
10 & 1000 & 250 & 2.5(3)\\
15 & 500 & 100 & 1.8(2)\\
15 & 500 & 250 & 1.1(1)\\
15 & 1000 & 100 & 2.0(2)\\
15 & 1000 & 250 & 1.2(2)\\
20 & 1000 & 100 & 1.7(3)\\
30 & 1000 & 100 & 1.1(1)\\
\hline
\end{tabular}
\caption{Values of the integrated auto-correlation time extracted with the $\Gamma$-method for several combinations of $L_d$, $\nstep$ and $\nrelax$, for the $N=21$ ensembles (left table) and the $N=41$ ensembles (right table). According to the results of Ref.~\cite{Bonanno:2018xtd}, where the joint dependence of the auto-correlation time of $\chi$ on $\beta$ and $N$ was studied using the same lattice volumes, lattice discretization and over-relaxation/heat-bath updating algorithms employed in the present investigation, for $N=21$ and $\beta=0.7$ we expect $\tauint^{(\mathrm{PBCs})}\sim 10^4-10^5$ standard updating steps using PBCs, while for $N=41$ and $\beta=0.65$ an even larger auto-correlation time. For a fair comparison, the quantity $\tauint^{(\mathrm{PBCs})}$ has to be compared with $\tauint \times (\nstep + \nrelax)$.}
\label{tab:tauint}
\end{table}

In the following, we will focus on the data obtained for combinations
of parameters $L_d$, $\nrelax$, $\nstep$ for which $\tauint$ is 
of order $1$, as such small values of $\tauint$ afford a reliable estimate
of the error on $a^2\chi$.
No investigation of the full dependence of $\tauint$ on
$\nstep$ will be attempted, as Tab.~\ref{tab:tauint} shows that its 
value does not seem to change appreciably, even when $\nstep$ is doubled.
Instead, we will focus on the regime $\nrelax \ge 50$, 
$\nstep \ge 500$, $L_d \geq 6$ for the $N=21$ ensemble 
and $L_d \geq 10$ for the $N=41$ ensemble.

The values of $\tauint$ are provided in Tab.~\ref{tab:betaL} for $N=21$ at different values of the coupling and of the lattice volume. 
Remarkably, no significant variation is observed in either $\tauint$ 
or $\hat{\ESS}$ as the coupling or the lattice volumes are varied at 
fixed $(L_d, \nstep, \nrelax)$. This is a further confirmation of the
previous observation that $\hat\ESS$ seems to only depend on $L_d$ and 
$\nstep$, and provides evidence of the robustness of 
the non-equilibrium method.

\begin{table}
\centering
\begin{tabular}{|ccccccc|}
\hline
$\beta$ & $L$ & $L_d$ & $\nstep$ & $\nrelax$ & $\tauint$ & $\hat\ESS$\\
\hline
 0.65 & 114 & 24 & 1000 & 50 & 0.52(4) & 0.72(1) \\
 0.7 & 114 & 24 & 1000 & 50 & 0.64(4) & 0.731(6) \\
 0.7 & 161 & 24 & 1000 & 50 & 0.68(5) & 0.73(1) \\
 0.75 & 114 & 24 & 1000 & 50 & 0.56(3) & 0.71(1) \\
\hline
\end{tabular}
\caption{Comparison of $\tauint$ and $\hat\ESS$ for different values of the coupling $\beta$ and the size of the lattice $L$}
\label{tab:betaL}
\end{table}

\subsection{Efficiency of the method}

\begin{figure}[!t]
\centering
\includegraphics[scale=0.45,keepaspectratio=true]{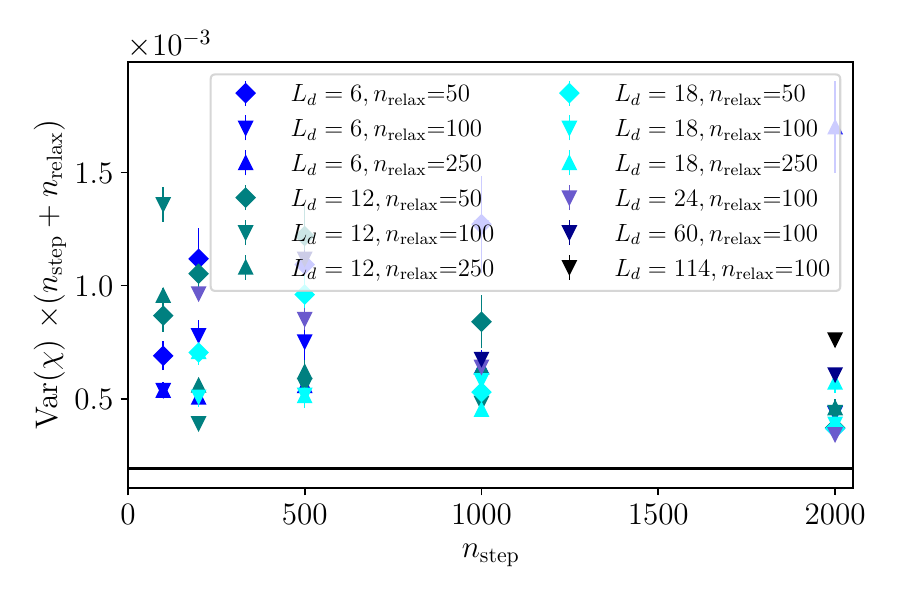}
\includegraphics[scale=0.45,keepaspectratio=true]{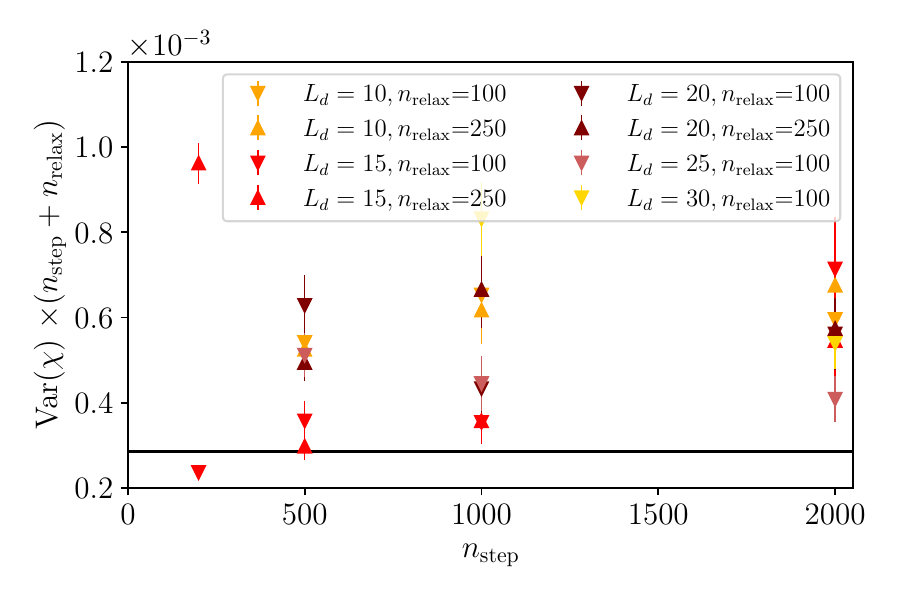}
\caption{Behavior of the variance of the topological susceptibility, multiplied by the cost per evolution (in units of the Monte Carlo update), as a function of $\nstep$ for several values of $\nrelax$ and $L_d$, for $N=21$, $\beta=0.7$ (left panel) and $N=41$, $\beta=0.65$ (right panel). The black line is the same quantity for the PTBC algorithm, from Refs.~\cite{Berni:2019bch,Bonanno:2022yjr}.}
\label{fig:effnstep}
\end{figure}

\begin{figure}[!t]
\centering
\includegraphics[scale=0.45,keepaspectratio=true]{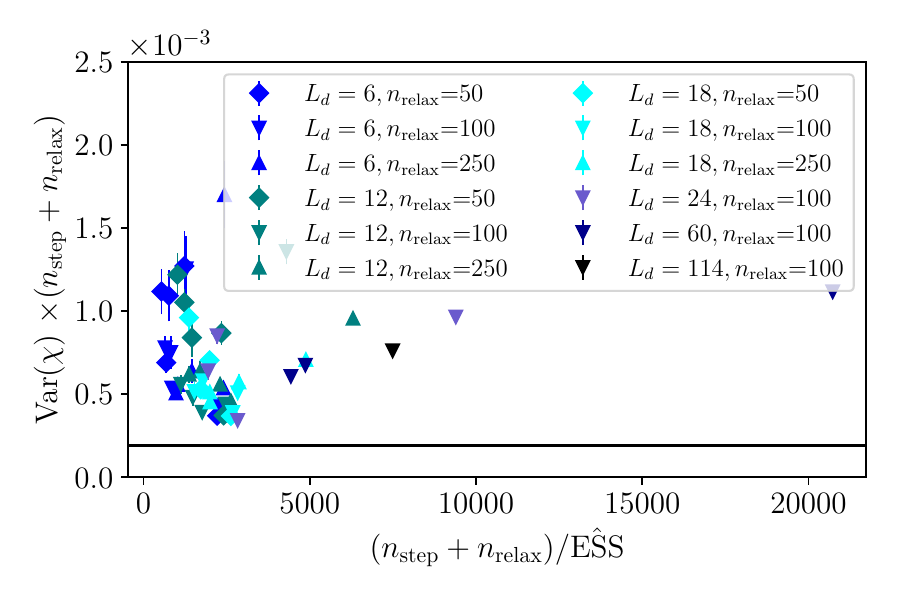}
\includegraphics[scale=0.45,keepaspectratio=true]{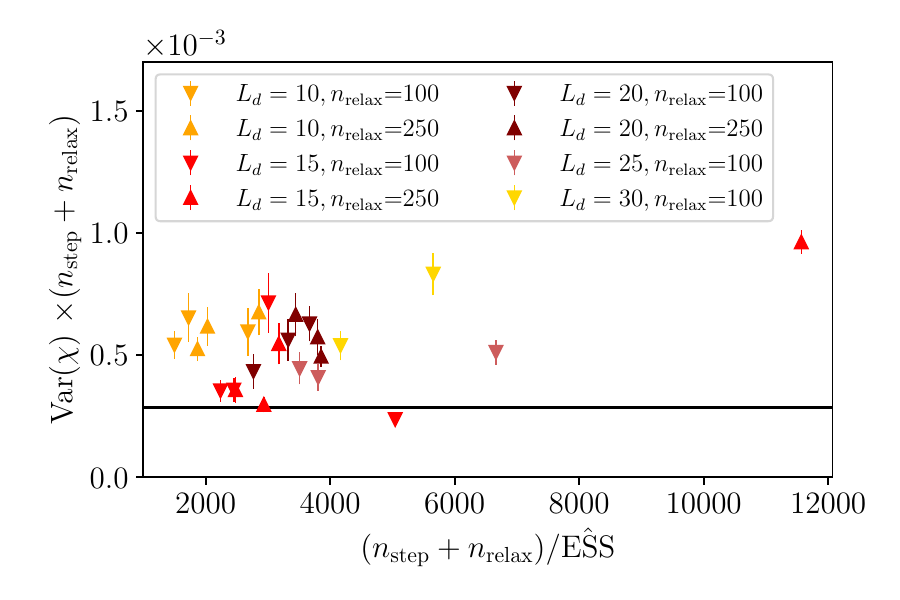}
\caption{Variance of the topological susceptibility, multiplied by the cost per evolution (in units of Monte Carlo updates), as a function of $(\nstep+\nrelax)/\hat{\ESS}$ for several values of $\nrelax$ and $L_d$, for $N=21$, $\beta=0.7$ (left panel) and $N=41$, $\beta=0.65$ (right panel). The black line is the same quantity for the PTBC algorithm, from Refs.~\cite{Berni:2019bch,Bonanno:2022yjr}.}
\label{fig:effess}
\end{figure}

In this section, we assess the efficiency of the non-equilibrium
method, using the product between the variance $\chi$ and the cost of each evolution,
$\Var(\chi)_{\NE}\times (\nstep+\nrelax)$,
as a figure of merit. 
Note that this is the quantity to minimize in order to reach the maximum efficiency in 
evaluating $\chi$, and this is the quantity on which a comparison with the
PTBC approach will be based.

The quantity $\Var(\chi)_{\NE}$ is computed directly from the sample of values of $\chi$
obtained \emph{after} reweighting according to 
Eq.~\eqref{eq:rew_like_jarz}. 
Some insight into this figure of merit can be gained 
from using Eq.~\eqref{eq:ESS_var}, and taking
into account the effects of auto-correlation. It then follows 
that
\beq\label{eq:variance}
\Var(\chi)_{\NE} \times (\nstep + \nrelax)  \simeq \Var(\chi)_p \frac{2 \tauint}{\hat\ESS} \times (\nstep + \nrelax).
\eeq
where $\Var(\chi)_p$ would be the variance of the topological susceptibility 
when sampled \emph{directly} from the target
distribution $p$ (i.e., with PBCs). Note that in principle, this is a quantity 
which is only dependent on the latter distribution, and independent from the
parameters of the non-equilibrium algorithm. 
Indeed, the effect of the auto-correlations
and of using Eq.~\eqref{eq:rew_like_jarz} to estimate observables are
accounted for, respectively, by the presence of $\tauint$ and $\hat\ESS$.

This figure of merit is displayed in Fig.~\ref{fig:effnstep} as a function
of $\nstep$, for several combinations of $L_d$ and $\nrelax$. 
The values obtained for the non-equilibrium estimations do not seem
to have any definite behaviour as $\nstep$ is increased. However, they cluster around
a value that is larger, but within $1.5\times 10^{-3}$, of those obtained with the 
PTBC approach, see Refs.~\cite{Berni:2019bch,Bonanno:2022yjr}.

Further insight into the efficiency of the non-equilibrium method can be 
gained by separating the effects of the reweighting step in
Eq.~\eqref{eq:rew_like_jarz} from the rest of the non-equilibrium procedure. 
This can be achieved simply by displaying the figure of merit in
Eq.~\eqref{eq:variance} as a function of $(\nstep+\nrelax)/\hat{\ESS}$.
This is done in Fig.~\ref{fig:effess}, from which it can be inferred that
$\Var(\chi)_{\NE}$ is not a function of $\hat{\ESS}$ alone.
The reason is twofold. First, the quantity we are using, $\hat{\ESS}$ is 
but an estimator of the ``true'' effective sample size. We refer to 
Appendix~\ref{sec:app_var} for further discussion. Second, a quantitative
characterization of the behaviour of $\tauint$ as a function of $\nrelax$,
$L_d$ and $\nstep$ is not attempted in this work. Although this would be needed
to unravel the full dependence of the figure of merit on the specific parameters
used in the non-equilibrium evolution, we believe that the above analysis
is a solid step in that direction, as it provides the region in which
the optimal values of $L_d$, $\nrelax$ and $\nstep$ may be found.

\section{Conclusions}\label{sec:conclu}

In this work, we have presented a first exploration of an non-equilibrium Monte
Carlo setup designed to mitigate the problem known as topological freezing.
We have carried out our analysis using the $2d$ $\CP^{N-1}$ model as test bed,
owing to their combination of numerical simplicity and physical non-triviality.
The new approach outlined in this manuscript has some features in common
with the PTBC algorithm originally proposed by M.~Hasenbusch. It consists 
in starting from a thermalized ensemble generated with OBCs, and gradually 
switching to PBCs along a non-equilibrium Monte Carlo evolution. 
At the end of the latter, expectation values with PBCs 
can be computed through a reweighting-like formula which is tightly related
to Jarzynski's equality.

This method is able to reproduce the expected value of the topological susceptibility
of the $\CP^{N-1}$ models. Moreover, it is possible to gauge the reliability
of the said reweighting-like formula using two different figures of merit:
the Effective Sample Size and the Kullback--Leibler divergence, which provide
consistent results. The efficiency of the method, quantified in terms of the
product of the variance of the final result with the computational effort of one measurement
is also shown to be comparable to the one found in the PTBC approach.

Put in perspective, the above study provides a broad framework for the
application of non-equilibrium methods to address the issue of critical 
slowing down in lattice simulations. It can be thought as a different kind
of Monte Carlo simulation, and could be implemented, for instance, by
varying the value of $\beta$ rather than of some parameter controlling the 
type of boundary conditions. In principle, one could start by
an equilibrium sampling of the configuration space of the system for values 
of the inverse coupling characterized by small 
auto-correlations. Then, the coupling could be gradually increased through 
non-equilibrium evolutions, to values at which the system would be strongly
auto-correlated at equilibrium. The meaning of $\nstep$ and $\nrelax$ would
then be the same as above, while $L_d$ would be replaced by a new parameter
$\Delta \beta$ describing the non-equilibrium changes in $\beta$. We highlight
that this approach has already been implemented in the $4d$ $\SU(3)$ pure-gauge
theory in Ref.~\cite{Caselle:2018kap}, although not with the aim of mitigating
the effects of critical slowing down.

Another relevant possible future application of the present
out-of-equilibrium method is represented by systems that include fermionic degrees
of freedom. This does not pose any additional theoretical difficulties, nor
does it contribute in principle with any computational overhead. Indeed, the calculation
of the action needed at each change of the parameter controlling the boundary conditions
for the gauge fields entering the fermion determinant is performed both at 
the beginning and at the end of each Hybrid Monte Carlo trajectory anyways. 
Alternatively, a prior distribution with periodic boundary conditions for 
the fermion determinant can be used, making the fermionic contribution to the
work exactly zero.

Finally, given the recently-established connection with Jarzynski's equality and the
theoretical framework of Stochastic Normalizing Flows (SNFs) in
Ref.~\cite{Caselle:2022acb}, our work sets the stage for an application of our
proposal to SNFs, where a stochastic part (which is given by the
out-of-equilibrium evolutions discussed in this study) is combined with the
discrete layers (parametrized by neural networks) that compose Normalizing
Flows. The training of such layers would of course need a possibly lengthy
procedure, which is however performed only once. Such an approach has the potential to greatly improve the
efficiency of the non-equilibrium evolutions as, in principle, a considerably lower
amount of Monte Carlo steps would be needed to achieve the same efficiency when
sampling the target distribution. Another natural future 
outlook of the present investigation is to implement the non-equilibrium setup
in a more physical and realistic model, such as the $4d$ $\SU(3)$ pure-gauge theory. 
We plan to investigate both ideas in the near future.

\acknowledgments
We thank A.~Bulgarelli, M.~Caselle, E.~Cellini and M.~Panero for insightful and helpful discussions. The work of C.~B.~is supported by the Spanish Research Agency (Agencia Estatal de Investigación) through the grant IFT Centro de Excelencia Severo Ochoa CEX2020-001007-S and, partially, by grant PID2021-127526NB-I00, both funded by MCIN/AEI/10.13039/ 501100011033. A.~N.~acknowledges support by the Simons Foundation grant 994300 (Simons Collaboration on Confinement and QCD Strings) and from the SFT Scientific Initiative of INFN. The work of D.~V.~is supported by STFC under Consolidated Grant No.~ST/X000680/1. The numerical simulations were run on machines of the Consorzio Interuniversitario per il Calcolo Automatico dell'Italia Nord Orientale (CINECA).

\appendix

\section*{Appendix}

\section{Evaluating $\hat{\ESS}$ as an estimator of the Effective Sample Size}
\label{sec:app_var}

\begin{figure}[!t]
\centering
\includegraphics[scale=0.75,keepaspectratio=true]{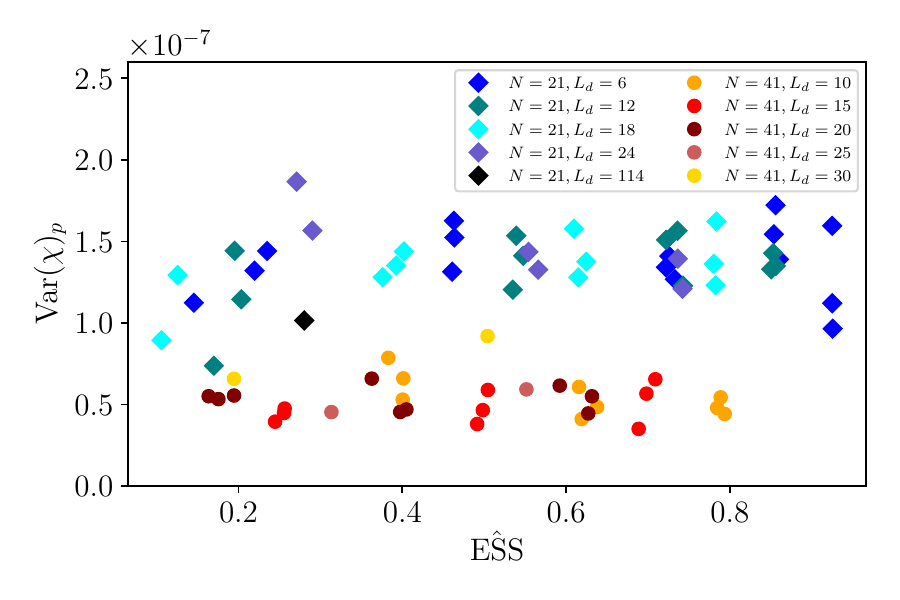}
\caption{Variance of the topological susceptibility sampled on PBCs, as a function of $\hat{\ESS}$ for several values of $\nstep$, $\nrelax$ and $L_d$, both for the $N=21$ and $N=41$ ensembles.}
\label{fig:varess}
\end{figure}

From Eq.~\eqref{eq:variance}, the variance $\Var(\chi)_p$ obtained in a system
with PBCs
can be related to the variance of the same quantity, but computed 
with non-equilibrium methods, as follows,
\beq\label{eq:variance2}
\Var(\chi)_p = \langle (a^2 \chi)^2 \rangle - \langle a^2 \chi \rangle^2 \simeq \Var(\chi)_{\NE} \frac{\hat{\ESS}}{2 \tauint}~.
\eeq
Now, as already stated in the main text, this quantity should be (approximately)
independent of the method used to compute $a^2\chi$ and of the
magnitude of the involved auto-correlations. In particular, this is
certainly valid if the ``true'' $\ESS$ is used in the right-hand side. We now
wish to check that this remains true also when $\ESS$ is replaced by its estimator
$\hat{\ESS}$.
If $\hat{\ESS}$ was a good estimator, then we would expect the left-hand side to be
independent of any change in the parameters of the non-equilibrium evolutions. 
In Fig.~\ref{fig:varess}, $\Var(\chi)_p$ is displayed as a function of $\hat{\ESS}$
for $N=21$ at $\beta=0.7$, and $N=41$ at $\beta=0.65$. The range of $\hat{\ESS}$
explored is evidence of the fact that the parameters that are being tuned do impact
on the non-equilibrium evolutions. Yet, for each value of $N$
separately, the values of $\Var(\chi)$
seem to cluster around a constant, with no discernible 
dependence on $\hat{\ESS}$. We thus conclude that
$\hat{\ESS}$ must be a good estimator of the true
effective sample size, at least in the range of parameters that
was explored.

\section{Relation between the Effective Sample Size and the Kullback--Leibler divergence}\label{app:DKL_vs_ESS}

\begin{figure}[!t]
\centering
\includegraphics[scale=0.48,keepaspectratio=true]{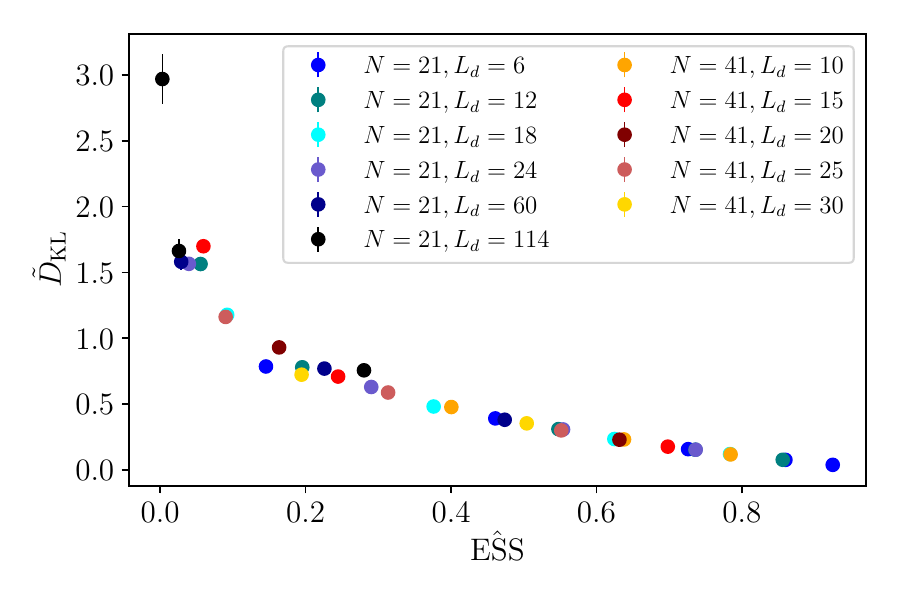}
\includegraphics[scale=0.48,keepaspectratio=true]{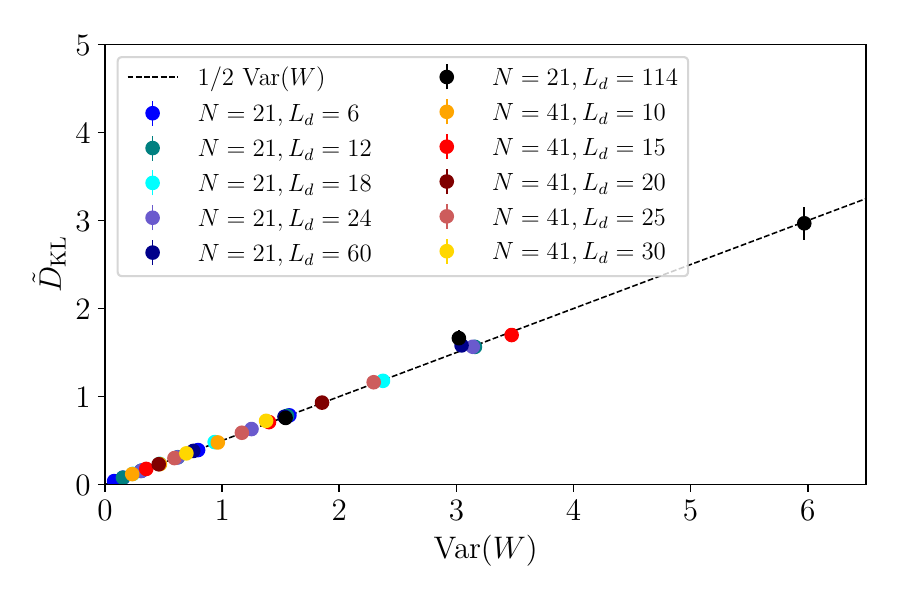}
\caption{Behavior of $\DKL$ as a function of $\hat{\ESS}$ (left panel) and as a function of the variance of the work $\Var(W)$ (right panel) for several values of the defect size $L_d$, both for the $N=21$ and $N=41$ ensembles.}
\label{fig:DKLESS_varw}
\end{figure}

In this study we considered two different figures of merit to quantify the
distance from equilibrium of our out-of-equilibrium evolutions: the
Kullback--Leibler divergence $\DKL$ of Eq.~\eqref{eq:DKL_therm_meaning} and the
estimator of the Effective Sample Size $\hat{\ESS}$ of Eq.~\eqref{eq:ESS_def}.
Given that the latter depends on the features of the probability distribution of
the weights $e^{-W}$ appearing in Eq.~\eqref{eq:rew_like_jarz}, and the former
on the probability distribution of the work $W$ itself, it is clear that these
two quantities are tightly related, although not in a straightforward way.
Indeed, our data seems to point out that $\DKL$ and $\hat{\ESS}$ are in a
one-to-one correspondence, with little to no dependence on the details of the
simulation (i.e., on $L_d$ and $\nstep$). In the left-hand
panel of Fig.~\ref{fig:DKLESS_varw}, $\DKL$ is displayed as a function of 
$\hat{\ESS}$. To an impressive degree of precision, the data points seem to
gather around the graph of an invertible function that relates $\hat{\ESS}$ to $\DKL$.
This signals that these two quantities are in a one-to-one correspondence.

Another aspect of the same tight relationship can be appreciated by
first highlighting a result from Ref.~\cite{Nicoli:2020njz}, 
\beq
\label{eq:dkl_varw}
\DKL \simeq \frac{1}{2} \Var(W),
\eeq
which was derived in the case of Normalizing Flows.
The above is strikingly demonstrated in left-hand panel of 
Fig.~\ref{fig:histograms}, where $\DKL$ is displayed as a function of $\Var(W)$.
The data seem to organize along a line with slope $1/2$, with remarkable 
precision, as expected from Eq.~\ref{eq:dkl_varw}.

\begin{figure}[!t]
\centering
\includegraphics[scale=0.75,keepaspectratio=true]{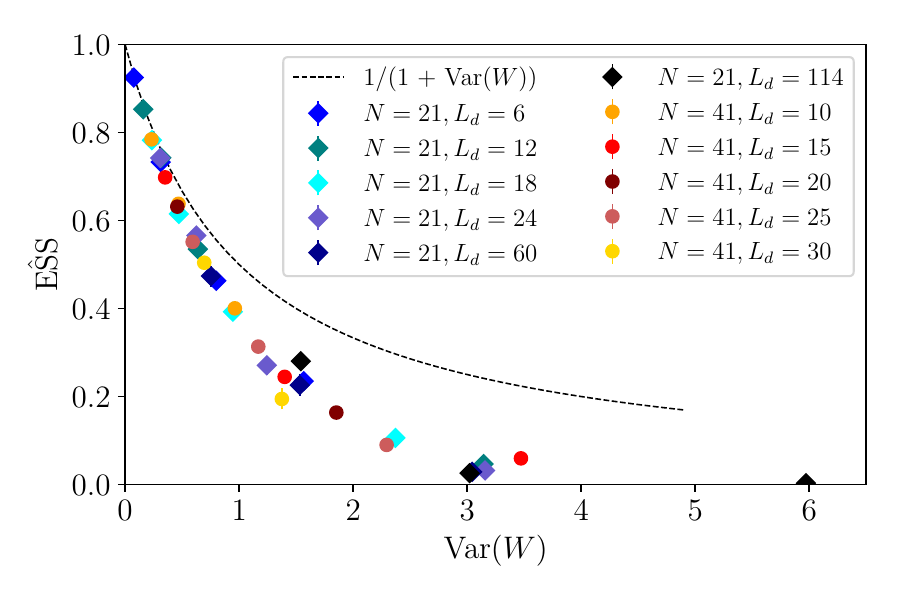}
\caption{Behavior of the estimator $\hat{\ESS}$ as a function of the variance of the work $\Var(W)$. The dashed line in the left panel represents our prediction from Eq.~\eqref{eq:ESS_varw} in the close-to-equilibrium regime.}
\label{fig:ESS_varw}
\end{figure}

Since the estimator of the Effective Sample Size is directly related to the
variance of the exponential average, see Eq.~\eqref{eq:ESS_expwvar}, it is
natural to think that it can be linked to the variance of the work itself. In
order to better understand this relationship, let us introduce the fluctuation
of the work $\delta W \equiv W - \braket{W}_{\mathrm{f}}$, where, by definition,
$\braket{\delta
W}_{\mathrm{f}} = 0$. It is straightforward to rewrite the $\hat{\ESS}$ as
follows,
\beq
\braket{e^{-2W}}_{\mathrm{f}} = e^{-2\braket{W}_{\mathrm{f}}} \braket{e^{-2 \delta W}}_{\mathrm{f}}, \quad \implies \quad \frac{1}{\hat{\ESS}} = e^{-2\DKL} \braket{e^{-2 \delta W}}_{\mathrm{f}}~.
\eeq
If our protocol is sufficiently close to equilibrium, we know from
Eq.~\eqref{eq:dkl_varw} that the work $W$ does not fluctuate much among
different non-equilibrium evolutions. Hence, if we assume that $\Var(W) \equiv
\braket{\delta W^2}_{\mathrm{f}} \ll 1$, then the typical value of $\delta
W$ along these evolutions is small. Thus, it is reasonable to perform the
following expansion,
\beq
\braket{e^{-2W}}_{\mathrm{f}} \simeq e^{-2\braket{W}_{\mathrm{f}}} (1 + 2 \, \Var(W)), \qquad \Var(W) \ll 1~.
\eeq
From Eq.~\eqref{eq:dkl_varw} we now obtain a simple relation between 
$\hat{\ESS}$ and the variance of the work in the close-to-equilibrium limit,
\beq\label{eq:ESS_varw}
\frac{1}{\hat{\ESS}} \simeq 1 + \, \Var(W), \qquad \Var(W) \ll 1~.
\eeq
The validity of this approximate relation is confirmed by our data, see
Fig.~\ref{fig:ESS_varw}, and is further evidence, at least in the regime 
of close-to-equilibrium evolutions, the estimator $\ESS$ is simply a function of $\Var(W)$.

\bibliographystyle{JHEP}
\bibliography{biblio}

\end{document}